\documentclass{aa}  
\usepackage{graphicx}
\usepackage{txfonts}
\usepackage{natbib}
\begin{document} 

   \title{Mass loss from inhomogeneous hot star winds}
   \subtitle{III. An effective-opacity formalism for line radiative
     transfer in accelerating, clumped two-component media, and
     first results on theory and diagnostics}

   \author{J.O. Sundqvist\inst{1}\and
     J. Puls\inst{1}\and
     S.P. Owocki\inst{2} }

   \institute{Universit\"atssternwarte M\"unchen, Scheinerstr. 1,
     81679 M\"unchen, Germany\\ \email{mail@jonsundqvist.com}\and
     University of Delaware, Bartol Research Institute, Newark,
     Delaware 19716, USA}
                                 
   \date{Received 2014-02-04; accepted 2014-05-27}

 
  \abstract
   {} 
   {To provide a fast and easy-to-use
   formalism for treating the reduction in effective opacity
   associated with optically thick clumps in an accelerating
   two-component medium.}
   {We develop and benchmark
   effective-opacity laws for continuum and line radiative transfer
   that bridge the limits of optically thin and thick clumps. We then
   use this formalism to i) design a simple method for
   modeling and analyzing UV wind resonance lines in hot, massive stars, and
   ii) derive simple correction factors to the line force driving the
   outflows of such stars.}
   {Using a vorosity-modified
   Sobolev with exact integration (vmSEI) method, we show that, for a
   given ionization factor, UV resonance doublets may be used to
   analytically predict the upward corrections in empirically inferred
   mass-loss rates associated with porosity in velocity space
   (a.k.a. velocity-porosity, or vorosity). However, we also show the
   presence of a solution degeneracy: in a two-component clumped wind
   with given inter-clump medium density, there are \textit{always}
   two different solutions producing the same synthetic doublet
   profile. We demonstrate this by application to SiIV and PV in B and
   O supergiants and derive, for an inter-clump density set to 1 \% of
   the mean density, \textit{upward empirical} mass-loss corrections
   of typically factors of either $\sim$\,5 or $\sim$\,50, depending
   on which of the two solutions is chosen. Overall, our results
   indicate that this solution dichotomy severely limits the use of UV
   resonance lines as direct mass-loss indicators in
   current diagnostic models of clumped hot stellar winds.

   We next apply the effective line-opacity formalism to the standard
   CAK theory of line-driven winds. A simple vorosity correction
   factor to the CAK line force is derived, which for normalized
   \textit{velocity} filling factor $f_{\rm vel}$ simply scales as
   $f_{\rm vel}^\alpha$, where $\alpha$ is the slope of the CAK
   line-strength distribution function. By analytic and numerical
   hydrodynamics calculations, we further show that in cases where
   vorosity is important at the critical point setting the
   mass-loss rate, the reduced line force leads to a \textit{lower
     theoretical} mass loss, by simply a factor $f_{\rm vel}$. On the
   other hand, if vorosity is important only above this critical
   point, the predicted mass loss is not affected, but the wind
   terminal speed is reduced, by a factor scaling as $f_{\rm
     vel}^{\alpha/(2-2\alpha)}$. This shows that porosity in velocity
   space can have a significant impact not only on the diagnostics,
   but also on the dynamics and theory of radiatively driven winds.}
{}
   \maketitle
%

\section{Introduction}
\label{intro}

It has been known for several years now, that the powerful
line-radiation driven winds of hot, massive stars are inhomogeneous
and highly structured on small spatial scales \citep[see overviews
  in][]{Puls08, Hamann08, Sundqvist12c}. Such \textit{wind clumping}
arises naturally from the strong line-deshadowing instability -- the
LDI, a fundamental property of line driving \citep[e.g.,][]{Owocki84,
  Owocki85} -- and affects both theoretical models and the diagnostic
radiative transfer tools needed to derive stellar and wind properties
from observed spectra of massive stars. For example, neglect of
clumping typically leads to observationally inferred mass-loss rates
that differ by more than an order of magnitude for the same star,
depending on which spectral diagnostic is used to estimate this mass
loss \citep{Fullerton06}.

Today's diagnostic wind models normally account for inhomogeneities by
simply assuming a one-component medium consisting of optically thin
clumps of a certain volume filling factor \citep[e.g.,][]{Hillier91,
  Puls06}. However, if individual clumps become optically thick, this
leads to an additional leakage of light -- not accounted for in the
filling factor approach -- through porous channels in between the
clumps. Such porosity can occur either in the second and third spatial
dimensions, or for spectral lines in \textit{velocity-space} due to
Doppler shifts in the rapidly accelerating wind (velocity-porosity, or
``vorosity'', Owocki 2008). In Papers I and II of this series
\citep{Sundqvist10, Sundqvist11}, we developed detailed
multi-dimensional wind models to study the effects of vorosity on the
formation of, in particular, the strong UV ``P-Cygni'' lines that are
the classical trademarks of massive star winds. A key general result
from these studies is that clumps indeed easily become optically thick
in such UV lines, and that the associated additional leakage of
photons leads to weaker line profiles than predicted by smooth or
volume filling factor models \citep[see also][]{Oskinova07, Hillier08,
  Surlan12, Surlan13}. But constructing realistic \textit{ab-initio}
radiation-hydrodynamic wind simulations that account naturally for
spatial and velocity-field porosity is an extremely challenging and
time-consuming task. Thus there is also a big need now for developing
simplified, paramterized models that can be more routinely applied to
diagnostic work on large samples of hot stars with winds, as well as
be used to investigate general effects on the theoretical predictions
and dynamics of such winds. This paper develops such a simplified
formalism, using effective quantities to simulate the reduction in
opacity associated with optically thick clumps. In contrast to our
earlier models \citep{Sundqvist10, Sundqvist11}, this ``effective
opacity''approach has the great advantage that it can be readily
implemented into the already existing NLTE (= non-local thermodynamic
equilibrium) atmospheric models normally used to analyze observed
spectra of hot stars with winds.

Sect. 2 develops and benchmarks effective-opacity laws to treat both
continuum and line radiative transfer in accelerating, stochastic
two-component media of (almost) arbitrary density contrasts. Sect. 3
then applies this effective opacity to line formation in hot stellar
winds, investigating the influence of velocity-porosity on UV line
diagnostics. Sect. 4 uses the same formalism to derive simple
correction factors of the effects of such vorosity on the line force
driving the outflows of hot, massive stars, providing simple scaling
relations for the effects on the global wind parameters mass-loss rate
and terminal speed. Finally, Sect.5 summarizes main results and gives
our conclusions.

\section{Basic effective-opacity formalism}
\label{theory} 

\begin{figure*}
  \begin{minipage}{6.0cm}
    \resizebox{\hsize}{!}
              {\includegraphics[angle=90]{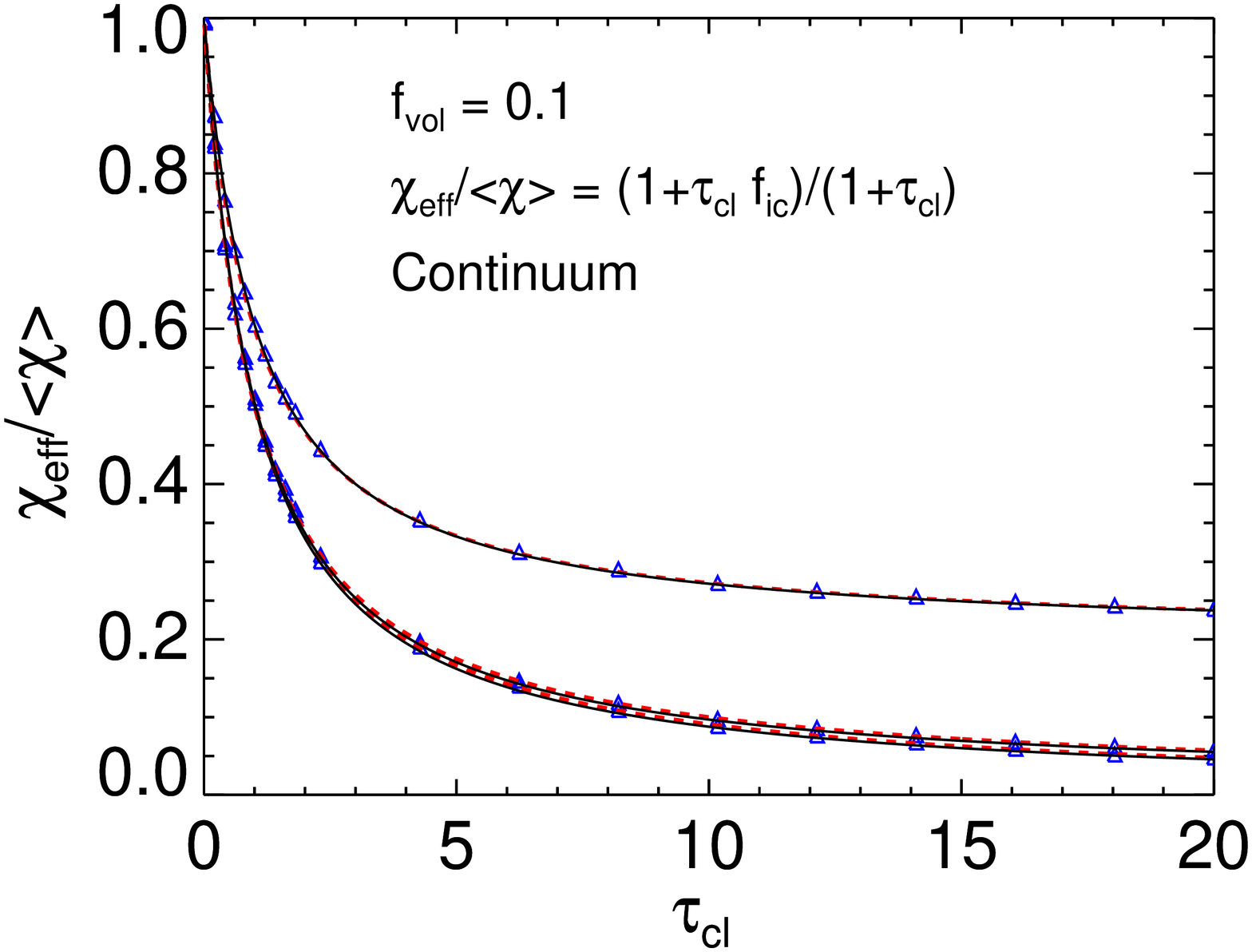}}
  \centering
   \end{minipage}
    \begin{minipage}{6.0cm}
    \resizebox{\hsize}{!}
            {\includegraphics[angle=90]{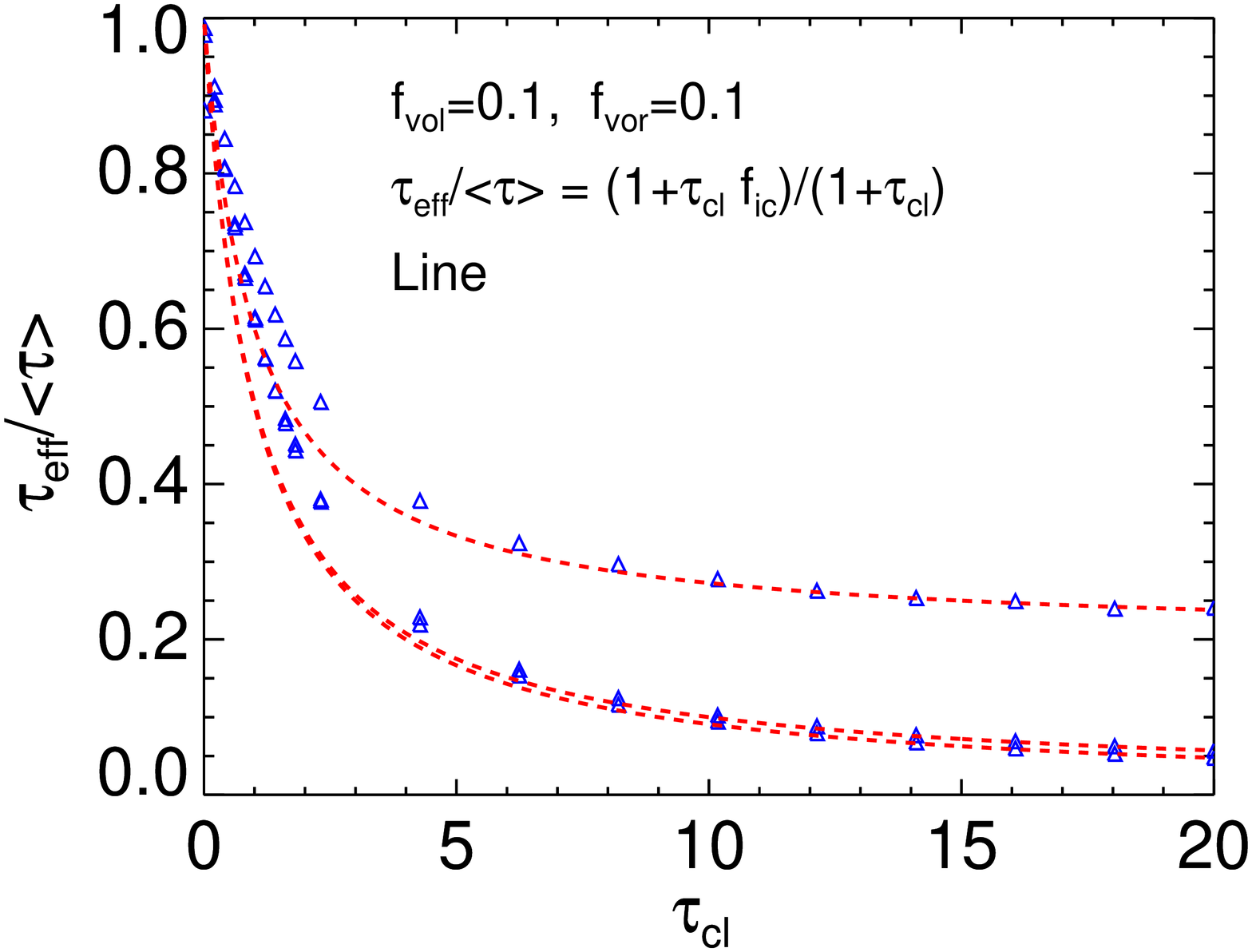}}
            \centering
   \end{minipage}         
    \begin{minipage}{6.0cm}
    \resizebox{\hsize}{!}
            {\includegraphics[angle=90]{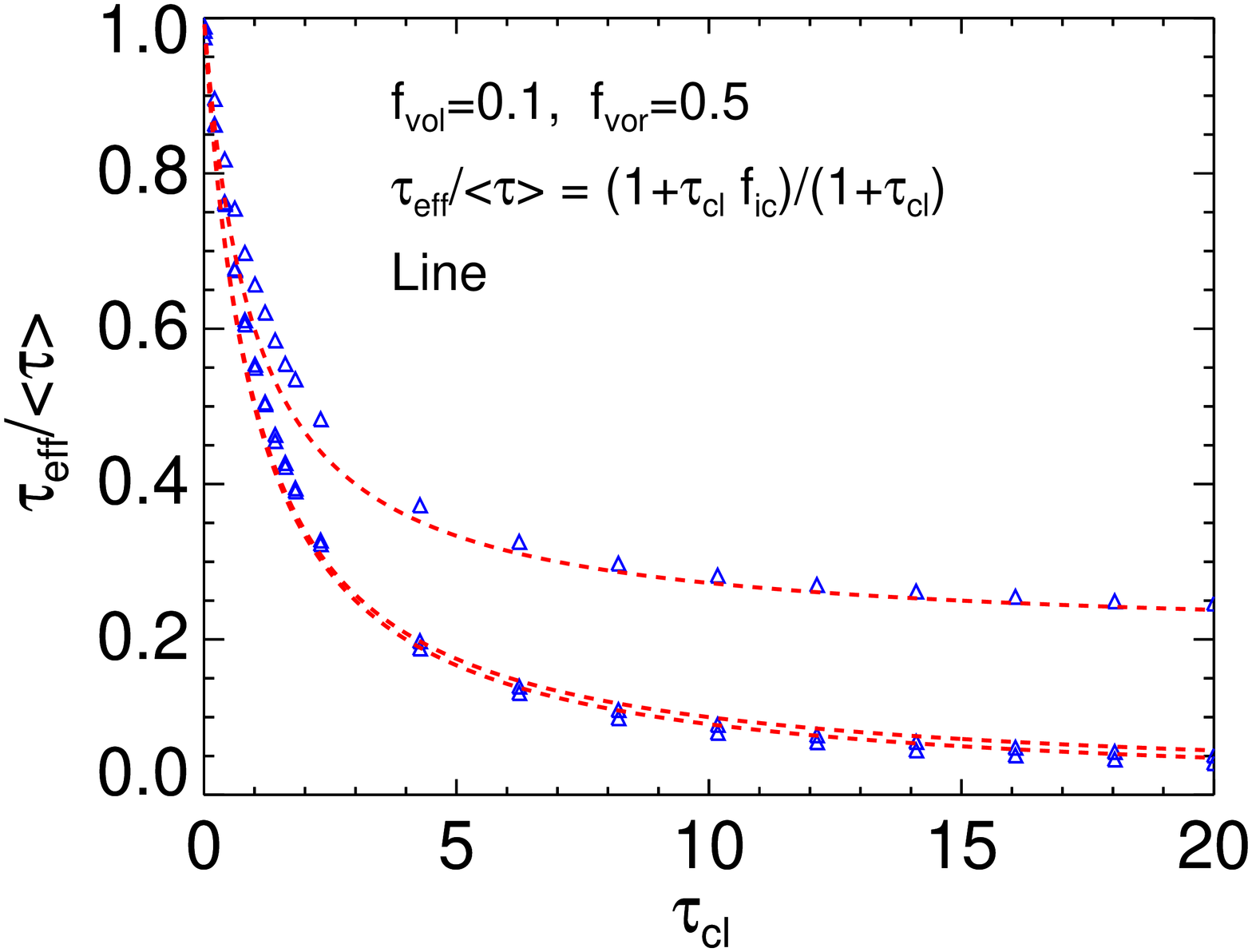}}
            \centering
   \end{minipage}         
  \caption{Ratios of effective to mean opacity/optical depth for
    continuum absorption (left) and line absorption (middle, right),
    as functions of clump optical depth and for the parameters given
    in the panels and inter-clump density parameters $f_{\rm ic} =$ 0,
    0.01, and 0.2 (where the $f_{\rm ic} = 0.2$ curve is the upper one
    in each panel). Note that the clump velocity span is $\delta \varv
    = \delta \varv_{\rm sm}$ in the middle panel and $\delta \varv = 5
    \delta \varv_{\rm sm}$ in the right one.  Red dashed lines show
    the effective opacity/optical depth laws indicated in the figure,
    blue triangles show results from the 3D-box experiments described
    in the text, and the black solid lines in the left panel show the
    analytic expression given in Appendix A.}
  \label{Fig:box_tests}
\end{figure*}

Although the primary aim of this paper is to develop a general and
useful formalism for \textit{line} opacity in accelerating, clumped
two-component media, it is instructive to first consider the
conceptually simpler case of \textit{continuum} opacity.  Below we
present full effective-opacity bridging laws for such continuum
absorption in a two-component medium with arbitrary
density-contrast. This extends our previously developed "porosity
models" \citep{Owocki04, Owocki06, Sundqvist12}, which have assumed
one of the components to be effectively void, and also provides
physical insights important for our following studies of line
absorption.

\subsection{Continuum absorption: A void interclump medium} 

Let us first very briefly review the case of an ensemble of clumps
embedded in an effectively void "inter-clump medium". For a given
clump optical depth $\tau_{\rm cl}$ and \textit{mean} free path
between clumps (a.k.a. the porosity length) $h$, the ``effective''
opacity (per unit length) is \citep[e.g.,][]{Feldmeier03, Owocki04}
\begin{equation} 
\chi_{\rm eff} = \frac{1-e^{-\tau_{\rm cl}}}{h}.   
\label{Eq:bridgetauc}
\end{equation} 
Here $\tau_{\rm cl} = \langle \chi \rangle h$, with mean opacity
$\langle \chi \rangle$ and porosity length $h \equiv l_{\rm cl}/f_{\rm
  vol}$, with $l_{\rm cl}$ the characteristic length scale of clumps
and $f_{\rm vol}$ the clump volume filling factor. Integrating the
clump interaction probability $P = 1-e^{-\tau_{\rm cl}}$ over an
exponential distribution in optical depths (or equivalently over
$l_{\rm cl}$, assuming a constant clump opacity),
\begin{equation}  
  f(\tau) = \frac{e^{-\tau/\tau_{\rm 0}}}{\tau_{\rm 0}}, 
  \label{Eq:dist_exp}
\end{equation}
with 
\begin{equation} 
  \langle \tau \rangle = \int_0^\infty \tau f(\tau) d\tau = \tau_0, 
\end{equation} 
then gives the ``inverse'' bridging law for the effective opacity
\citep{Sundqvist12},
\begin{equation} 
  \frac{\chi_{\rm eff}}{\langle \chi \rangle} = \frac{1}{1+\tau_{\rm 0}}
  = \frac{1}{1+\tau_{\rm cl}}.  
  \label{Eq:b_c_void}
\end{equation} 
In this equation, $\langle \chi \rangle$ is the mean opacity
calculated from a smooth model, or a structured model assuming
optically thin clumps, and $\tau_0 = \tau_{\rm cl} = \ \langle \chi
\rangle h$ now represents the \textit{mean} clump optical depth.  In
this paper, we assume this $\tau_{\rm cl}$ is statistically
isotropic. This results in an isotropic effective opacity
\citep{Sundqvist12}, as favored by recent empirical investigations of
X-ray line profile shapes in O-star winds \citep{Leutenegger13}. Note
that the porosity-associated reduction in effective opacity in this
clump+void continuum model thus depends only on porosity length $h$.

\subsection{Continuum absorption: Full bridging law for two-component media}

Eqn.~\ref{Eq:b_c_void} above neglects absorption in the inter-clump
(ic) medium. This is most probably a good assumption for continuum
radiative transfer in stellar winds with clumping properties set by
the LDI, but may be questionable for situations in which deep-seated
atmospheric clumping might be expected, for example envelope inflation
of stars that approach the Eddington limit \citep{Grafener12,
  Grafener13}, or porosity-mediated continuum-driven mass loss in such
stars \citep{Owocki04}. We therefore next generalize the porosity
model above to consider also the case where both components may
contribute to the total opacity. For mean density $\langle \rho
\rangle = f_{\rm vol} \rho_{\rm cl} + (1-f_{\rm vol}) \rho_{\rm ic}$,
we approximate the opacity in such general two-component media with
(see Appendix A)

\begin{equation}
  \frac{\chi_{\rm eff}}{\langle \chi \rangle} = 
  \frac{1 + \tau_{\rm cl} f_{\rm ic}}{1+\tau_{\rm cl}},
  \label{Eq:b_c_gen}
\end{equation}
where $f_{\rm ic} \equiv \rho_{\rm ic} / \langle \rho \rangle$ denotes
the contrast between inter-clump and mean density. Because of 
mass conservation, the
clump optical depth for mass absorption coefficient $\kappa$ and mean
opacity $\langle \chi \rangle = \kappa \langle \rho \rangle$ now
formally is
\begin{equation} 
  \tau_{\rm cl} = \rho_{\rm cl} \kappa l_{\rm cl} =
  \langle \chi \rangle h \ \big(1-(1-f_{\rm vol})f_{\rm ic}\big), 
\end{equation} 
however in most cases of interest the correction factor $1-(1-f_{\rm
  vol})f_{\rm ic}$ will be near unity, so that $\langle \rho \rangle
\approx \rho_{\rm cl} f_{\rm vol}$ and $\tau_{\rm cl} \approx \langle
\chi \rangle h$ still are good approximations. We further note in this
context that for processes with a density-independent mass absorption
coefficient $\kappa$ (like bound-free absorption of X-rays), the mean
opacity $\langle \chi \rangle = \kappa \langle \rho \rangle$ is not
directly affected by the presence of optically \textit{thin} clumps,
whereas for processes with $\kappa \propto \rho$ (like thermal
free-free emission), this mean opacity is enhanced by a clumping
factor $f_{\rm cl} \equiv \langle \rho^2 \rangle/\langle \rho
\rangle^2 \approx 1/f_{\rm vol}$ (where the last approximation assumes
a negligible contribution from the inter-clump medium) as compared to
a homogeneous model.

To verify the bridging-law ansatz eqn.~\ref{Eq:b_c_gen}, we compare to
an \textit{exact} expression (derived analytically by
\citealt{Pomraning91}, see also \citealt{Levermore86}) for the
emergent intensity in stochastic two-component media with spatially
constant opacities and with length scales of the individual components
distributed exponentially\footnote{Typically, such media are referred
  to as \textit{Markovian binary mixtures}, see \citet{Pomraning91}.},
as in our eqn.~\ref{Eq:dist_exp}. Appendix A gives the formidable
expression for this emergent intensity, along with a translation of
the parameters used by Pomraning et al. to those used in this
paper. Extensive testing comparing the analytic result with the simple
bridging law eqn.~\ref{Eq:b_c_gen} shows excellent agreement for a
broad range of conditions, as illustrated by the opacity curves as
functions of clump optical depth in the left hand panel of
Fig.~\ref{Fig:box_tests} (in which we also plot comparisons using a
3D-box model, as described in the following subsection). We note in
particular how previously known cases all are correctly reproduced by
the new bridging law:
\begin{itemize}
	\item  optically thin clumps: $\tau_{\rm cl} << 1 \ \rightarrow
	\ \chi_{\rm eff} = \langle \chi \rangle$
	\item two equal components: $f_{\rm ic}=
	1 \ \rightarrow \ \chi_{\rm eff} = \langle \chi
	\rangle$ 
	\item a negligible inter-clump medium density: \\
	 $f_{\rm ic} = 0 \rightarrow \chi_{\rm eff} = \langle \chi \rangle/(1+ \tau_{\rm cl})$ 
	 (eqn.~\ref{Eq:b_c_void}). 
\end{itemize} 

Another interesting limit to examine is that of optically thick clumps
and a tenuous but non-negligible inter-clump medium. In this case the
product-term in eqn~\ref{Eq:b_c_gen} decouples to yield $\chi_{\rm
  eff}/\langle \chi \rangle \approx 1/\tau_{\rm cl} + f_{\rm
  ic}$. This simple expression illustrates explicitly how in the case
of such black clumps the inter-clump medium may be viewed as gradually
filling in the porous channels between the clumps. In contrast to the
clump+void model, in which the opacity \textit{itself} saturates at
$\chi_{\rm eff} = 1/h$ (and thus becomes independent of the mean
opacity), in this general two-component model the \textit{ratio}
between the effective and mean opacities saturates, at $\chi_{\rm
  eff}/\langle \chi \rangle = f_{\rm ic}$ \citep[see
  also][]{Owocki04}. This means that, independent of the size of $h$,
the medium can always become optically thick provided the mean opacity
is high enough. Fig.~\ref{Fig:box_tests} demonstrates that the
opacity-ratio curve assuming a larger density contrast $f_{\rm ic} =
0.01$ is almost indistinguishable from that assuming a void
inter-clump medium, whereas the $f_{\rm ic} = 0.2$ curve indeed
approaches this $\chi_{\rm eff} / \langle \chi \rangle \approx f_{\rm
  ic}$ limit for very optically thick clumps.

\subsection{Line absorption in rapidly accelerating media}
\label{eff_line}

\begin{figure}
  {\includegraphics[angle=90,width=7.5cm]{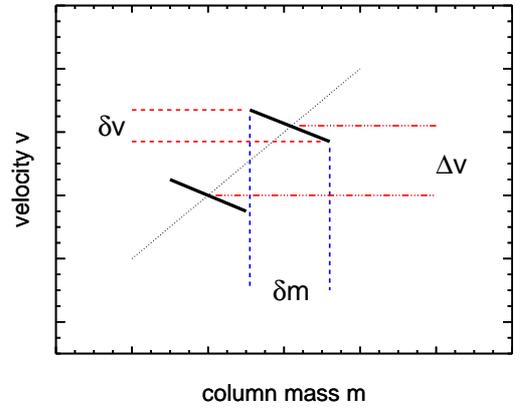}}
  \centering
  \caption{Sketch of the vorosity effect in a model with clump column
    mass $\delta m$, clump velocity span $\delta \varv$, and mean
    velocity separation between clumps $\Delta \varv$. In the case of
    an inter-clump medium with negligible mass, all line photons with
    resonance frequencies that do not coincide with any velocities
    $\delta \varv$ covered by the clumps will escape the wind without
    ever interacting with it, through the porous channels in velocity
    space set by $f_{\rm vor} \equiv \delta \varv/\Delta \varv$. Note
    that in the special case of the radially compressed model
    discussed in text, $\delta m = \delta m_{\rm sm}$, where $\delta
    m_{\rm sm}$ is the column mass contained within $\Delta \varv$ in
    the initially smooth wind, which has here been swept up in the
    clump.}
  \label{Fig:sketch}
\end{figure}

Due to the Doppler effect, the formation of spectral lines in a
clumped, accelerating medium differs conceptually from the continuum
case studied above. Namely, for example in the supersonic, rapidly
accelerating outflows of hot stars, each line photon can only interact
with the wind material within a very narrow spatial range, set by the
Sobolev length $L_{\rm Sob} = \varv_{\rm th}/\varv'$ for thermal speed
$\varv_{\rm th}$ and velocity gradient $\varv'$. The small extent of
this resonance zone makes it possible for line photons to leak through
the wind via porous channels in velocity space, without ever
interacting with the clumps. Hence we dub such leakage of light in
velocity space \textit{velocity-porosity}, or vorosity.

\paragraph{The clump optical depth in a spectral line.} 
The difference between spatial porosity and velocity-porosity becomes
particularly evident through the calculation of $\tau_{\rm cl}$. As
for the continuum case, we in this paper take this line clump optical
depth to be isotropic. Assuming then that the velocity span $\delta
\varv$ of individual clumps is greater than a few thermal widths
($\varv_{\rm th} \approx$\,5-10 km/s for a metal ion in a hot star
wind), the clump optical depth of a spectral line, normalized in terms
of the radial Sobolev optical depth $\tau_{\rm Sob}$ in a smooth (sm)
model (or one with optically thin clumps, if $\kappa \sim \rho$) can
be written as
\begin{equation}
  \frac{\tau_{\rm cl}}{\tau_{\rm Sob}} = \frac{\delta m}{\delta m_{\rm sm}}
    \left|\frac{\delta \varv_{\rm sm}}{\delta \varv}\right|,
    \label{Eq:taucl_mc}
\end{equation} 
where the $\delta m$'s and $\delta \varv$'s are the column masses and
velocity spans of the clump and and the smooth wind, respectively. Note
here that $\delta m_{\rm sm}$ and $\delta \varv_{\rm sm}$ as defined
in eqn.~\ref{Eq:taucl_mc} must be computed from the same length scale,
but that the ratio $\delta \varv_{\rm sm}/\delta m_{\rm sm}$ (and thus
the ratio $\tau_{\rm cl}/\tau_{\rm Sob}$) is independent of the choice
of this scale.

In a first general model, we now \textit{choose} $\delta \varv_{\rm
  sm}$ to be the velocity span a clump \textit{would have} if it
followed the smooth wind velocity law (so that $\delta m/\delta m_{\rm
  sm} = \rho_{\rm cl} l_{\rm cl}/(\langle \rho \rangle l_{\rm cl}) =
\rho_{\rm cl}/\langle \rho \rangle$), eqn.~\ref{Eq:taucl_mc} can be
further expressed as \citep[see also][]{Sundqvist10,Sundqvist11}
\begin{equation}
  \frac{\tau_{\rm cl}}{\tau_{\rm Sob}} = \frac{1}{f_{\rm vol} \ |\delta
    \varv/\delta \varv_{\rm sm}|} \big(1-(1-f_{\rm vol})f_{\rm ic}\big) \approx
  \frac{1}{f_{\rm vol} \ |\delta \varv/\delta \varv_{\rm sm}|} =
  \frac{1}{f_{\rm vor}},
  \label{Eq:taucl_line}
\end{equation}
where the second expression here again neglects the small correction
factor due to the inter-clump medium, and the third expression
introduces the velocity clumping factor $f_{\rm vor}$. This velocity
clumping factor is defined as the ratio of the velocity span of clumps
to their mean velocity separation $\Delta \varv$ (see
Fig.~\ref{Fig:sketch}, and also \citealt{Owocki08}),
\begin{equation}
  f_{\rm vor} \equiv \left| \frac{\delta \varv}{\Delta \varv} \right|,  
  \label{Eq:fvel} 
\end{equation} 
which is obtained in eqn.~\ref{Eq:taucl_line} by assuming clumps that,
on average, have their central positions distributed in velocity-space
according to the smooth wind expansion rate. In this case $\Delta
\varv \approx \varv_{\rm sm}' h$, using the definition of the porosity
length $h \equiv l_{\rm cl}/f_{\rm vol}$ as the \textit{mean} free
path between clumps, which then results in $f_{\rm vol} \ | \delta
\varv/\delta \varv_{\rm sm}| = (l_{\rm cl}/h)| \delta
\varv/(\varv'_{\rm sm} l_{\rm cl})| = | \delta \varv / \Delta \varv | =
f_{\rm vor}$. 

While eqn.~\ref{Eq:taucl_line} is a quite general expression for
$\tau_{\rm cl}$, a conceptually better understanding of how $f_{\rm
  vor}$ determines the velocity-porosity effect can be obtained by
considering a simple model of radially compressed clumps consisting of
swept up material from an initially smooth wind (similar to
predictions of present-day LDI simulations). For this model then (see
sketch in Fig.~\ref{Fig:sketch}), the initially smooth wind contained
within $\Delta \varv$ will have a column mass $\Delta m = \delta
m_{\rm sm} = \delta m$, so that
\begin{equation}
  \frac{\tau_{\rm cl}}{\tau_{\rm Sob}} = \frac{\delta m}{\delta m_{\rm sm}}
    \left|\frac{\delta \varv_{\rm sm}}{\delta \varv}\right| 
    = \left|\frac{\Delta \varv}{\delta \varv}\right| 
    = \frac{1}{f_{\rm vor}}. 
\end{equation} 
Moreover, we can of course also in this model choose to instead define
the smooth wind quantities on the clump length scale, so that in this
case $\delta \varv = (\delta \varv/\delta \varv_{\rm sm}) \varv_{\rm
  sm}' l_{\rm cl}$ and, since the mean free path between clumps in
Fig.~\ref{Fig:sketch} still is $h = \Delta \varv / \varv_{\rm sm}'$,
the velocity clumping factor $f_{\rm vor} = f_{\rm vol} \ |\delta
\varv/ \delta \varv_{\rm sm}|$. This shows that this one-dimensional
radially compressed model is fully consistent with the more general
eqn.~\ref{Eq:taucl_line}. Note also that $f_{\rm vor}$ as defined here
is an \textit{un-normalized} quantity, which only for the case that
the velocity field inside the clumps follows the smooth wind velocity
law becomes a normalized quantity.

In summary, the key point is that the final expression for the clump
optical depth depends \textit{only} on this velocity clumping factor
(and a correction factor for the mass contained in the inter-clump
medium).

\paragraph{Effective opacity bridging law for spectral lines.}  
For line clump optical depth $\tau_{\rm cl}=\tau_{\rm Sob}/f_{\rm
  vor} \times (1-f_{\rm ic}(1-f_{\rm vol})) \approx \tau_{\rm Sob}/f_{\rm
  vor}$, we now (in analogy with the continuum case) suggest to write
the vorosity-modified effective line opacity as
\begin{equation}
  \frac{\chi_{\rm eff}}{\langle \chi \rangle} = \frac{1 + \tau_{\rm
      cl} f_{\rm ic}}{1+\tau_{\rm cl}},  
  \label{Eq:b_l_gen}
\end{equation}
where the correction factor $\tau_{\rm cl} f_{\rm ic}$ in the
nominator now essentially assumes that the inter-clump medium, on
average, follows the smooth wind expansion rate, and so fills in all
holes in velocity space not covered by the dense clumps. This simple
treatment of the inter-clump medium allows us to account for the fact
that UV line profiles in dense O-star winds often exhibit zero
residual flux (in particular at high velocities), which is direct
observational evidence that at least some material must be present at
a wide range of wind velocities \citep[e.g.,][]{Sundqvist10}.  Another
inherent assumption in eqn.~\ref{Eq:b_l_gen}, retained throughout this
paper, is that the ionization states of the two components of the
medium can be approximated with one ``effective" state (for a first
attempt to build a two-component model that relaxes this, see
\citealt{Zsargo08}).

Eqn.~\ref{Eq:b_l_gen}, like the continuum opacity bridging law, gives
the expected results in previously studied limits, namely:
%
\begin{itemize}
	\item optically thin clumps: $\tau_{\rm cl} \ll 1 \ \rightarrow
	\ \chi_{\rm eff} = \langle \chi \rangle$
	\item two equal components: $f_{\rm ic}=
	1 \ \rightarrow \ \chi_{\rm eff} = \langle \chi \rangle$   
	\item optically thick clumps and a negligible inter-clump
          medium: \\  $f_{\rm ic}=0 \ \ \tau_{\rm cl} \gg 1
	 \ \rightarrow \tau_{\rm eff} = \chi_{\rm eff}/\varv_{\rm sm}' 
         = f_{\rm vor}$,           
\end{itemize} 
where the last limit is valid for rays in the radial direction and
illustrates how we perform the final radiative transfer calculations
on a background smooth model, which is the simplifying key point in
developing this kind of effective-opacity formalisms. The limit shows
further how in the absence of an in-filling inter-clump medium, the
effective line optical depth in the radial direction saturates at a
value given simply by $f_{\rm vor}$, which means that the escape
fraction of line photons in this case is $e^{-f_{\rm
    vor}}$. Physically, this is analogous to the escape fraction of
light in traditional radiative transfer models, but with the spatial
holes between randomly distributed atomic absorbers replaced here by
velocity holes between clumps that are randomly distributed about
their mean in velocity space (see also discussion in
\citealt{Sundqvist11}, their Appendix A). Since $f_{\rm vor} =1$ still
gives an escape fraction $e^{-1}$, this property shows the importance
of defining $f_{\rm vor}$ as an \textit{un-normalized} quantity (not
bound between 0 and 1) in such effective-opacity models, so that
optically thick conditions with $f_{\rm vor} \rightarrow \infty$ can
be reached.

It is, finally, important to realize that although the proposed
continuum and line effective-opacity laws have the same principal
forms, the conceptual difference between spatial porosity and
velocity-porosity is reflected in the calculation of the clump optical
depth, which for continuum opacity depends on porosity length $h$, and
for line opacity on velocity clumping factor $f_{\rm vor}$.

\paragraph{Benchmarking with 3D-box experiments.}
Before considering the specific case of line formation in rapidly
accelerating stellar winds, we first test the general validity of the
new bridging law. For this purpose, we randomly distribute clumps in a
3D-box according to (constant) pre-specified volume filling factors,
clump length scales, and opacities for both clumps and the inter-clump
medium. For each such randomization, a ray is fired from the bottom of
the box and the emergent intensity $I$ computed at the top. This
procedure is then repeated until statistical errors in the averaged
intensities are sufficiently small, resulting in a final effective
optical depth $\tau_{\rm eff} = - \ln \langle I \rangle$. For clump
length scales distributed exponentially according to
eqn.~\ref{Eq:dist_exp}, we have verified that for continuum opacity
this set-up gives perfect agreement with the analytic expression
provided in Appendix A (and thus also with the continuum
effective-opacity bridging law), as illustrated by the blue triangles
in the left hand panel of Fig.~\ref{Fig:box_tests}.

To study line formation, we add to the continuum set-up a simple
velocity field proportional to height $Z$ in the box, and evaluate the
Sobolev optical depth at the points where a given line-frequency has
been Doppler shifted into resonance (which can be more than one due to
the random distribution of clumps). We assign an inter-clump optical
depth $\tau_{\rm ic}$ only to rays that do not intersect any clump.
This effectively assumes an inter-clump medium density $\rho_{\rm
  ic}/\langle \rho \rangle \la 0.2$, so that clumps still dominate the
total absorption whenever they are intersected. To facilitate
comparisons, we re-formulate eqn.~\ref{Eq:b_l_gen} in terms of the
vorosity-modified effective optical depth\footnote{The mean optical
  depth in the random-box experiment can be computed from the specific
  probabilities that a clump or the inter-clump medium is hit. After
  integrating over an exponential distribution in clump length scales,
  this results in $\langle \tau \rangle = f_{\rm vor}/(1+f_{\rm vor})
  \tau_{\rm cl} + 1/(1+f_{\rm vor})\tau_{\rm ic}$. Accounting
  additionally for the fact that more than one clump can be hit due to
  overlaps in velocity space, we approximate $\langle \tau \rangle
  \approx f_{\rm vor}\tau_{\rm cl} + 1/(1+f_{\rm vor})\tau_{\rm ic} =
  \tau_{\rm Sob}(1+ f_{\rm ic}/(1+f_{\rm vor})) \approx \tau_{\rm
    Sob}$ for the $f_{\rm ic} \la 0.2$ considered here.}
\begin{equation}
  \frac{\tau_{\rm eff}}{\langle \tau \rangle} \approx 
  \frac{\tau_{\rm eff}}{\tau_{\rm Sob}} =  
  \frac{1 + \tau_{\rm cl}f_{\rm ic}}{1+\tau_{\rm cl}}.    
  \label{Eq:b_l_tau}
\end{equation}
%
The middle and right hand panels of Fig.~\ref{Fig:box_tests}
compare effective optical depth curves of the line-opacity bridging
law to the 3D box simulations, and show an overall good
agreement. Whereas the middle panel assumes a velocity span of clumps
$\delta \varv$ that follows the underlying mean expansion $\delta
\varv_{\rm sm}$, the right panel shows that the bridging law is valid
also for velocity spans that deviate from this expansion rate, here
using $\delta \varv = 5 \delta \varv_{\rm sm}$; since the opacity
curves in the two panels are plotted against the clump optical depth
$\propto \tau_{\rm Sob}/f_{\rm vor}$, they automatically adjust for
the different assumed velocity filling factors and so appear similar
in the figure. As in the continuum case, the figure shows that the
effective optical depth approaches $\langle \tau_{\rm eff} \rangle /
\tau_{\rm Sob} \approx f_{\rm ic}$ for very optically thick clumps.
In summary, these 3D box experiments thus provide good general support
for the proposed effective-opacity bridging law to treat line opacity
in a rapidly accelerating, stochastic two-component medium.

Backed up by these results, we next consider two applications of the
effective-opacity bridging law developed in this section, namely: i) a
very simple method for computing and analyzing UV resonance lines from
the winds of hot, massive stars, and ii) an equally simple method for
estimating the vorosity effect on the driving line force of such
winds.


\section{Application I: Line diagnostics}
\label{diagnostics} 

The standard Sobolev with exact integration \citep[SEI,][]{Lamers87}
method for computing UV wind resonance lines uses the Sobolev
approximation to first obtain the source function, after which the
formal integral of radiative transfer is solved exactly to compute the
emergent flux spectrum. This section first develops a simple
vorosity-modified SEI method (vmSEI), by using the effective-opacity
bridging law introduced above, and then demonstrates how it may be
analytically applied to obtain vorosity corrections for empirically
inferred mass-loss rates.

\subsection{A vorosity-modified SEI method}

\begin{figure*}
  \begin{minipage}{4.55cm}
    \resizebox{\hsize}{!}
              {\includegraphics[angle=90]{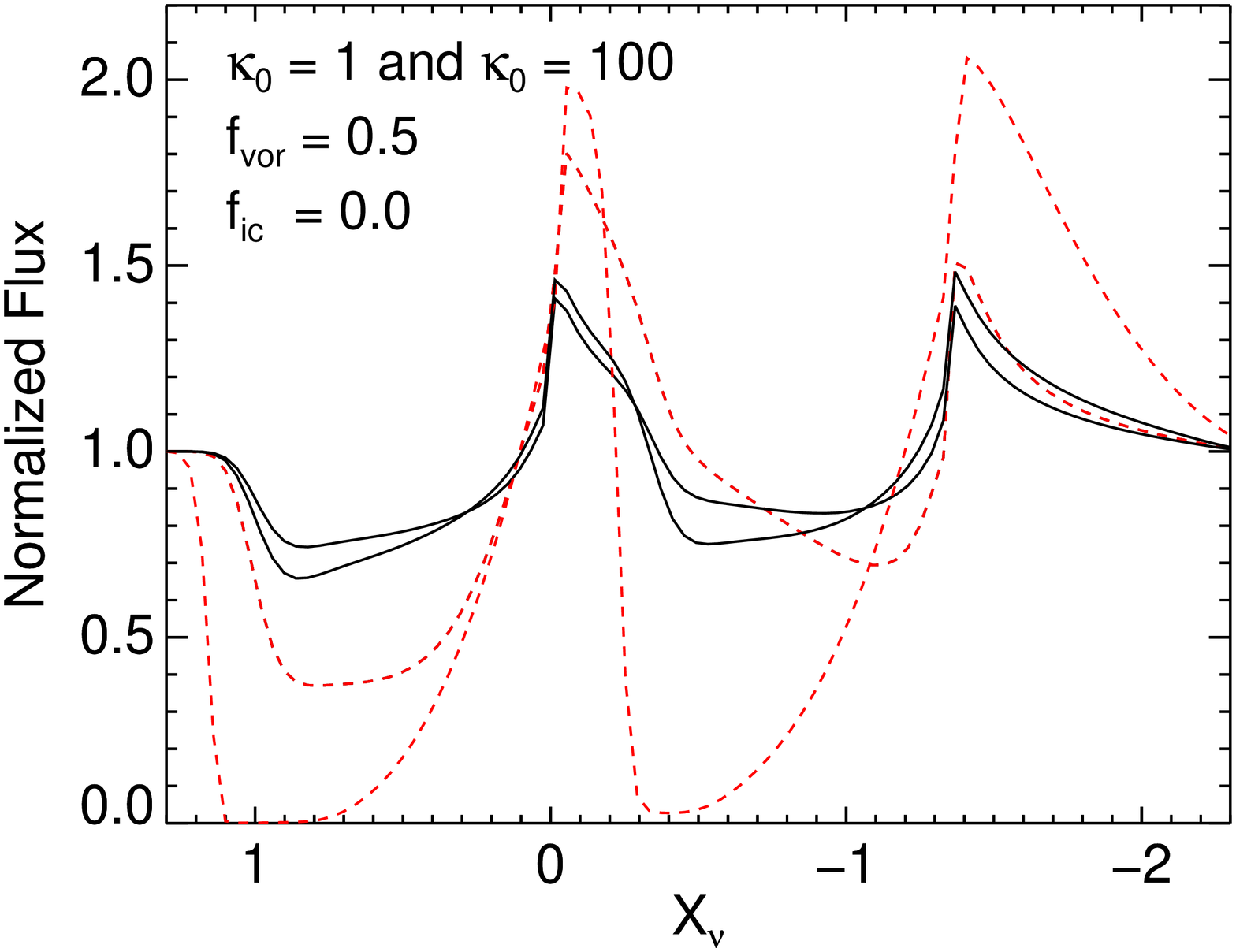}}
  \centering
   \end{minipage}
    \begin{minipage}{4.55cm}
    \resizebox{\hsize}{!}
            {\includegraphics[angle=90]{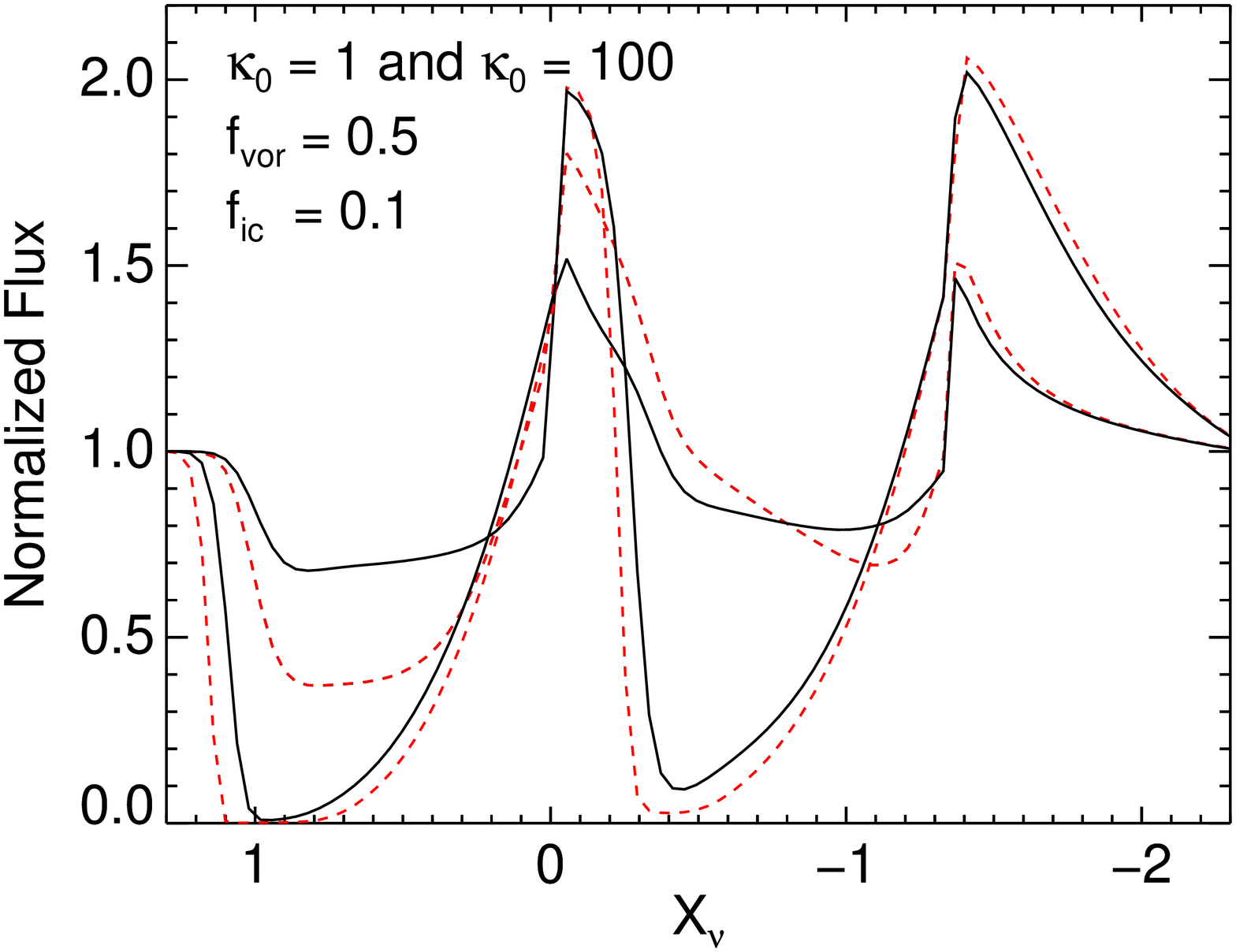}}
            \centering
   \end{minipage}         
    \begin{minipage}{4.55cm}
    \resizebox{\hsize}{!}
            {\includegraphics[angle=90]{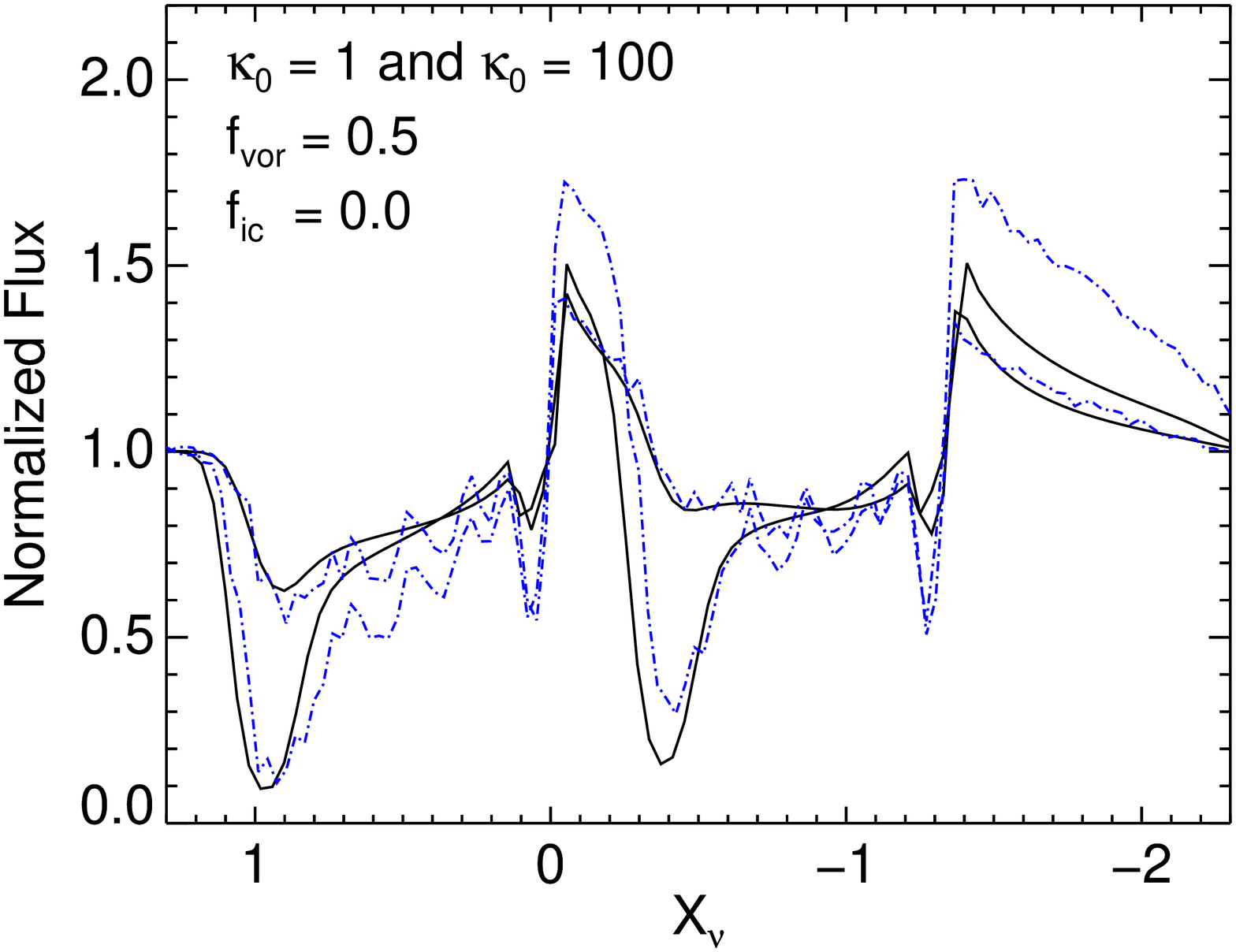}}
            \centering
   \end{minipage}         
    \begin{minipage}{4.55cm}
    \resizebox{\hsize}{!}
            {\includegraphics[angle=90]{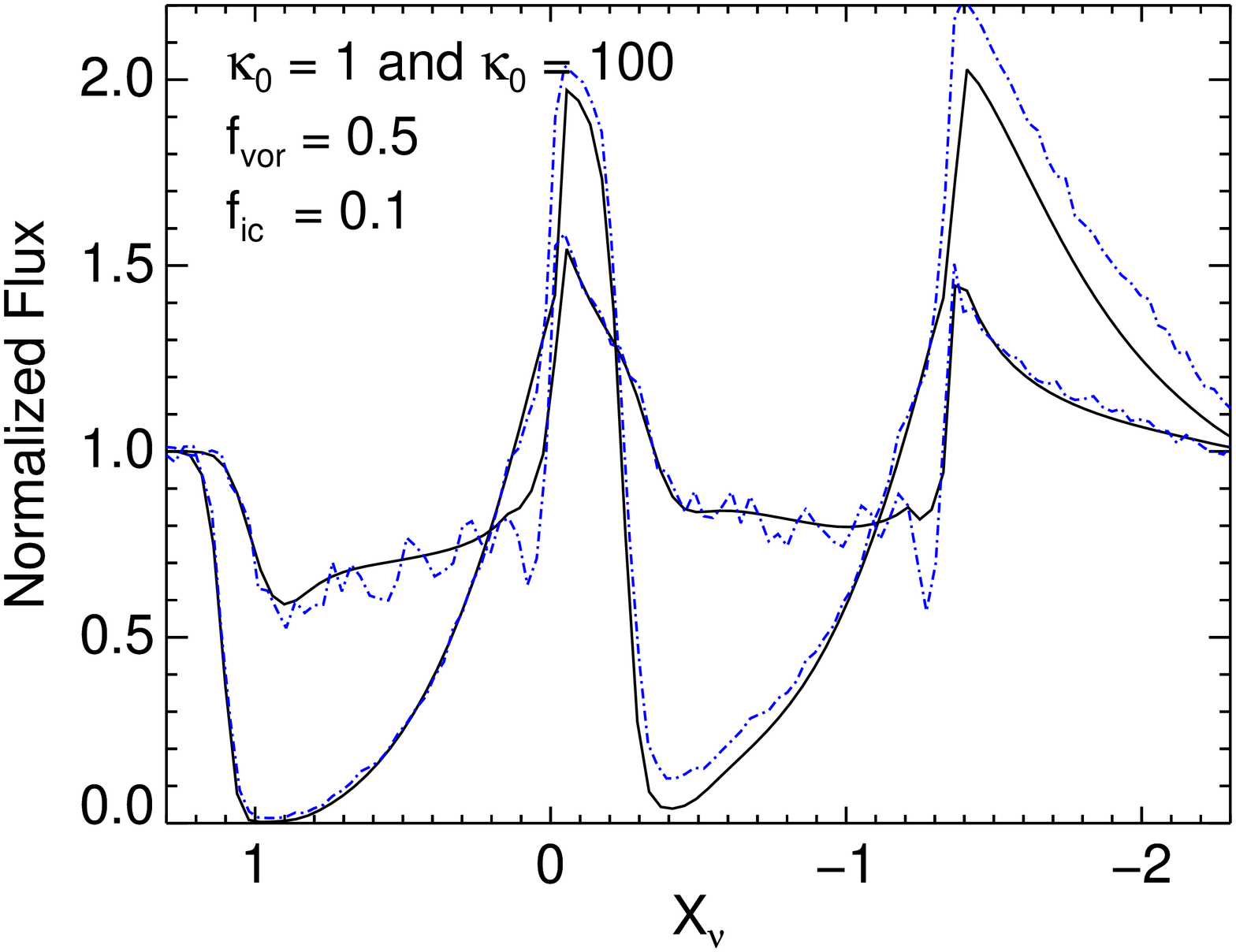}}
            \centering
   \end{minipage}         
  \caption{Synthetic UV line profiles in units of normalized frequency
    $X_\nu = (\nu/\nu_0-1)c/\varv_\infty$, and for velocity separation
    $\varv_{\rm sep}/\varv_{\infty} = 1.35$ between the line-centers
    of the two doublet components (which corresponds to the separation
    of the PV resonance doublet in a wind with $\varv_{\infty} =
    2000$\,km/s). For the line-strengths and clumping parameters
    indicated in the panels and described in table~\ref{Tab:params}
    and in text, the black lines show profiles computed using the
    vmSEI method developed here, the red dashed lines (two leftmost
    panels) show results from the standard SEI method (assuming a
    smooth wind), and the blue dashed lines (two rightmost panels)
    show results from the Monte-Carlo method by
    \citet{Sundqvist10}. Note that the velocity clumping factors
    indicated in the panels are based on the original definition
    eqn.~\ref{Eq:fvel} rather than the modified
    eqn.~\ref{Eq:fvel_eff}.}
  \label{Fig:vmSEI_prof}
\end{figure*}

The opacity of a trace element in a UV wind resonance line can, for
mass-loss rate $\dot{M}$ and wind terminal speed $\varv_\infty$, ion
fraction $q$ of the considered element $i$, and abundance with respect
to hydrogen $\alpha_{\rm i} = n_{\rm i}/n_{\rm H}$, be conveniently
expressed as a dimensionless opacity-parameter \citep{Hamann81}
\begin{equation} 
  \kappa_0 = \frac{\dot{M} q}{R_\star \varv_\infty^2} \frac{\pi
    e^2/m_{\rm e}c}{4 \pi m_{\rm H}} \frac{\alpha_{\rm i}}{1+4Y_{\rm
      He}} f_{\rm lu} \lambda_0,
  \label{Eq:kappa0_tot}
\end{equation} 
where it has been assumed that the entire ion population resides in
the ground state (normally a safe assumption for the lines considered
in this paper, e.g. \citealt{Puls08}). In eqn.~\ref{Eq:kappa0_tot},
$R_\star$ is the stellar radius, $Y_{\rm He}$ the helium number
abundance, $f_{\rm lu}$ the oscillator strength of the transition, and
$\lambda_0$ the rest wavelength. With this parametrization, the radial
Sobolev optical depth in a smooth wind becomes
 \begin{equation}
  \tau_{\rm Sob}(r) = \frac{\kappa_{\rm 0}}{r^2 w {\rm d} w/{\rm d}r},
  \label{Eq:tausob_app}
\end{equation}
where $r$ is measured in units of $R_{\star}$ and $w =
\varv/\varv_{\infty}$. To account for the effects of optically thick
clumping, we now simply replace the opacity parameter $\kappa_0$ with
the radius dependent (but angle independent) effective opacity
\begin{equation}
  \kappa_{\rm eff}(r) = \left( \frac{1 + \tau_{\rm
      cl} f_{\rm ic}}{1+\tau_{\rm cl}} \right) \kappa_0 .  
  \label{Eq:eff_kap}
\end{equation}
For pre-specified velocity clumping factor $f_{\rm vor}$ and
inter-clump medium density parameter $f_{\rm ic}$, implementing
eqn.~\ref{Eq:eff_kap} in a SEI code thus reduces to evaluating the
clump optical depth $\tau_{\rm cl}(r) = \tau_{\rm Sob}/f_{\rm vor}
\times (1-(1-f_{\rm vol})f_{\rm ic})\approx \tau_{\rm Sob}/f_{\rm
  vor}$ at each radial grid point. Using the obtained effective
opacity the source functions are then calculated from the angle
dependent effective Sobolev optical depth $\tau_{\rm eff}^{\rm
  Sob}(r,\mu) = \kappa_{\rm eff}/(r^2 w Q)$, with $Q \equiv \mu^2dw/dr
+ (1-\mu^2)w/r$ for directional cosine $\mu$, and the formal integral
finally solved following the standard SEI approach.

For a modestly overlapping resonance doublet and a standard wind
velocity law\footnote{0.99 corresponds to $\varv_{\rm min} = 0.01
  \varv_\infty$ for a $\beta=1$ velocity field.} $w =
(1-0.99/r)^\beta$, here with $\beta = 1$, the two leftmost panels of
Fig.~\ref{Fig:vmSEI_prof} compare line profiles computed using this
vmSEI approach with profiles computed using the standard SEI model,
for intermediate and strong lines with $\kappa_0 = 1$ and 100. The
figure shows clearly the basic velocity-porosity effect, namely weaker
line-profiles for a given line-strength parameter $\kappa_0$,
consistent with all previous work on the effects of optically thick
clumps on UV wind lines \citep{Oskinova07, Sundqvist10, Sundqvist11,
  Surlan12, Surlan13}.

Moreover, for the models with void inter-clump medium the absorption
in the lines saturates at a level above zero, at $\sim e^{-f_{\rm
    vor}}$ (as discussed in Sect.~\ref{eff_line}), whereas in the case
of $f_{\rm ic} = 0.1$ the strong $\kappa_0 = 100$ line recovers the
absorption blackness of the smooth model. Another important result of
vorosity evident from the figure, is that the relative strength
between the blue and red components can differ significantly from the
expected factor of 2 in optical depth \citep[stemming from the
  oscillator strength ratio, see also][]{Prinja10, Sundqvist11,
  Prinja13};
this last property is examined in detail in Sect.~\ref{Bsuper}.

\subsection{Comparison to Monte-Carlo simulations}

\begin{table*}
\begin{minipage}{\textwidth}
    \centering
    \caption{Input parameters in the Monte-Carlo, multi-dimensional, stochastic wind simulations by \citet{Sundqvist10, Sundqvist11}.}
        \begin{tabular}{ l l l l l l }
        \hline \hline 
        Parameter    & Clump volume & Clump velocity & Porosity & Inter-clump & Shock jump \\
                     & filling factor & span & length & density parameter & velocity \\
        Symbol      & $f_{\rm vol}$ & $| \delta \varv / \delta \varv_{\rm sm}|$ & $h$  & $f_{\rm ic}$ & $\varv_{\rm j}$   \\
                \hline
        \end{tabular}
    \label{Tab:params}
\end{minipage}
\end{table*}

We next compare this new vmSEI model to profiles computed using an
extension (to treat doublets) of the method developed by
\citet{Sundqvist10}. This creates a multi-dimensional stochastic wind
by taking 1-D snapshots and phasing them randomly in patches of a
parameterized angular size, here 3 degrees, and then computes
synthetic spectra via a Monte-Carlo radiative transfer
technique. These stochastic wind models are created such that they
preserve the basic properties of LDI simulations, while still allowing
for different quantitative wind structure properties by the adjustment
of a number of input parameters , as given by Table 1. In the test
cases displayed in Fig.~\ref{Fig:vmSEI_prof}, we have assumed a
``velocity-stretch'' porosity law \citep[e.g.,][]{Sundqvist12}
$h/R_\star = w$, a clump onset radius $r_{\rm cl} = 1.1 R_\star$, and
shock jump velocity $\varv_{\rm j} = 0.1 w$, where the shock jump
velocity in the vmSEI model is simulated using the same ``turbulent''
velocity parameter $\varv_{\rm turb}$ as in the traditional SEI
approach. The rightmost two panels of Fig.~\ref{Fig:vmSEI_prof} show
an overall good agreement between the two methods, providing general
support for the usage of the much simpler effective-opacity method for
the quantitative analysis of hot stellar wind spectra.

One feature in the stochastic models not captured by the vmSEI
approach is the redward excess emission in strong lines. This excess
is caused by photon trapping within the resonance zones and by
increased back-scattering due to multiple such resonance zones
\citep{Lucy84, Puls93, Sundqvist10}, which allows light to escape
primarily when emitted on the red side of the line profile; such
multi-scattering effects cannot be simulated within the simple
effective opacity method developed in this paper, but does not affect
the \textit{absorption} line strength that is the primary focus here
(and in general when using unsaturated resonance lines as diagnostic
tools). Moreover, in particular the $\kappa_0 = 100$ line with a void
inter-clump medium shows a prominent absorption-dip towards the blue
edge of the profile \citep[see also][]{Sundqvist10, Surlan12}. In the
stochastic wind models, overlapping clumps in velocity space and the
finite extent of the line profile lead to an increase in $f_{\rm vor}$
at high velocities, and so results in more efficient absorption in the
outermost wind than in the accelerating parts of it. To account for
this absorption effect in the vmSEI model, we re-write the velocity
clumping factor as \citep[][their Appendix A]{Sundqvist11}
\begin{equation} 
f_{\rm vor} \approx \left| \frac{\delta \varv + \varv_{\rm th}}{\Delta
  \varv} \right| = f_{\rm vol}\left| \frac{\delta \varv}{\delta \varv_{\rm
    sm}}\right| + \frac{L_{\rm Sob}}{h}.
\label{Eq:fvel_eff}
\end{equation} 
This equation shows that, for a given inter-clump density, the other
vorosity-related input parameters in the stochastic wind models (see
Table 1) can be combined into one, the velocity clumping factor
$f_{\rm vor}$. It also demonstrates how the absorption-dips in the
stochastic models result from the very large Sobolev lengths in the
outermost wind, which give the radiative transfer a pseudo-continuum
character that reduces the velocity-porosity effect (which requires
rapid acceleration). In this respect, we note also that the
Monte-Carlo calculations performed by \citet{Sundqvist11} directly on
LDI simulations confirm that the dip is present also in such
hydrodynamical wind models calculated from first principles.

It is important to realize here, however, that using $f_{\rm vor}$ as
an empirical input-parameter when calculating the effective opacity,
or as a fit parameter when modeling observed line profiles,
\textit{automatically} accounts for this outer-wind absorption effect
(if present).

We next demonstrate how, indeed, unsaturated resonance doublets can be
used to directly diagnose the radial behavior of the velocity filling
factor, as well as to derive vorosity corrections for empirically
inferred mass-loss rates.


\subsection{An analytic method for vorosity mass-loss corrections}
\label{Bsuper}

\begin{figure}
              {\includegraphics[angle=90,width=6.5cm]{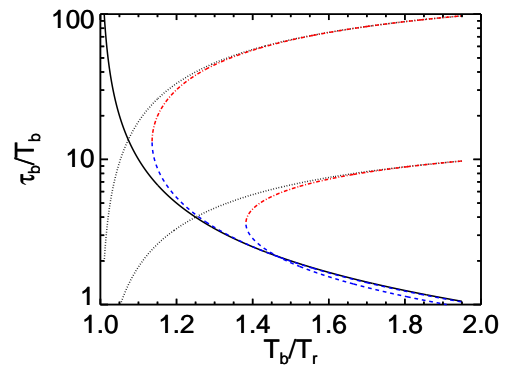}}
  \centering
  \caption{vmSEI optical depth correction factors $ \tau_{\rm
      b}/T_{\rm b}$ as function of fiducial resonance doublet optical
    depth ratio $T_{\rm b}/T_{\rm r}$. The black solid and dotted
    lines are computed using eqns.~\ref{Eq:mdot_void} and
    \ref{Eq:mdot_icm}, and the blue dashed and red dashed-dotted lines
    show the two solution branches from solving eqn.\ref{Eq:mdot_full}
    for inter-clump density parameters $f_{\rm ic} = 0.01$ and 0.1.}
  \label{Fig:mdot_theor}
\end{figure}

\begin{figure}
              {\includegraphics[angle=90,width=6.5cm]{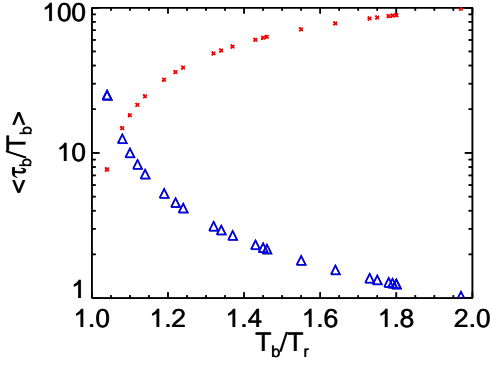}}
  \centering
  \caption{The two solution branches for vmSEI mass loss times ion
    correction factors for the B supergiant sample of \citet{Prinja10}
    and inter-clump density parameter $f_{\rm ic} = 0.01$ (see text).}
  \label{Fig:mdot_bsuper}
\end{figure}

As first pointed out in the context of hot star winds by
\citet{Massa08}, in a homogeneous (or optically thin clumped) wind the
optical depths of the individual components of a resonance doublet
must be set by the ratio of their oscillator strengths. On the other
hand, in a wind consisting of some mixture of thin and thick clumps
(and a potentially non-void inter-clump medium), this is no longer
necessarily the case. By allowing the oscillator strength ratio in the
standard SEI model to be a free parameter \citep[or alternatively
  examining only completely separated doublets, see][]{Prinja10,
  Prinja13}, we can empirically infer ``fiducial'' Sobolev optical
depths $T_{\rm b}$ and $T_{\rm r}$ for the blue and red components. We
show below how these fiducial optical depths derived from a
\textit{smooth} wind model then can be used to analytically obtain
$f_{\rm vor}$ and correction factors to the product of mass-loss rate
and ion fraction.

\paragraph{A void inter-clump medium.} Neglecting first absorption within a
tenuous inter-clump medium, we demand that $T_{\rm b}$ and $T_{\rm r}$
be equal to the vorosity-modified effective optical depths,
\begin{equation} 
  T_{\rm b} = \tau_{\rm eff}^{\rm b} =  \frac{\tau^{\rm b}}{1+\tau^{\rm b}/f_{\rm vor}}, 
\end{equation}
\begin{equation} 
  T_{\rm r} = \tau_{\rm eff}^{\rm r} =  \frac{\tau^{\rm b}/2}{1+\tau^{\rm b}/(2 f_{\rm vor})},
\end{equation}
where for simplicity we have denoted the Sobolev optical depth by
$\tau$, and, of course, in the physics-based vorosity model the
\textit{real} oscillator strength ratio $f_{\rm b}/f_{\rm r}$ must be
used; to allow for an analytic investigation, we here assume $f_{\rm
  b}/f_{\rm r} =2$, which is true for most UV resonance lines of
interest (for example PV, CIV, SiIV). For given $T_{\rm r}$ and
$T_{\rm b}$ then, we thus have two equations and two unknowns and can
solve immediately for the velocity clumping factor and the vorosity
correction in optical depth,
\begin{equation} 
  f_{\rm vor} = T_{\rm b} \frac{1}{2-T_{\rm b}/T_{\rm r}},   
   \label{Eq:fvel_void}
\end{equation} 
\begin{equation} 
  \frac{\tau^{\rm b}}{T_{\rm b}}  =
   \frac{(\dot{M} q)_{\rm vmSEI}}{(\dot{M} q)_{\rm SEI}} = \frac{1}{T_{\rm b}/T_{\rm r}-1}. 
   \label{Eq:mdot_void}
\end{equation}
All quantities in these equations are \textit{local}, i.e., by
scanning the observed line profile one finds $T_{\rm b}$ and $T_{\rm
  r}$ as a function of radius and so also $f_{\rm vor}(r)$ and the
optical depth correction $(\tau_{\rm b}/T_{\rm b})(r)$.

Analysis of the remarkably simple eqns.~\ref{Eq:fvel_void} and
\ref{Eq:mdot_void} shows that in the physical limit $T_{\rm b}/T_{\rm
  r} \rightarrow$ 2, the vorosity mass-loss (times ion fraction)
correction goes to unity and $f_{\rm vor} \rightarrow \infty$; this is
expected since in this case the smooth wind conditions should be
recovered, and in this clump+void model the only way to achieve this
is by making all clumps optically thin, i.e. $\tau_{\rm cl} =
\tau/f_{\rm vor} \rightarrow 0$, which requires $f_{\rm vor}
\rightarrow \infty$.  In the opposite limit $T_{\rm b}/T_{\rm r}
\rightarrow 1$, the mass-loss correction approaches $\infty$ and
$f_{\rm vor} \rightarrow T_{\rm b}$; such huge mass-loss correction
for this case is also as expected, since we showed in
Sect.~\ref{eff_line} that in this limit $\tau_{\rm eff} = f_{\rm
  vor}$, and the profile-strength then becomes \textit{independent} of
mass loss.

We note also that while the velocity clumping factor depends on the
actual empirically inferred profile strength $T_{\rm b}$, the quantity
$\tau_{\rm b}/T_{\rm b}$ depends \textit{only} on the ratio $T_{\rm
  b}/T_{\rm r}$. This very simple property makes it particularly
appealing to consider the behavior of such vorosity corrections for
mass loss, which we indeed focus on in the analysis below.

\paragraph{Adding an absorbing inter-clump medium.} 

Adding a non-void inter-clump medium that contributes to the total opacity 
gives for the equations to be solved, 
\begin{equation} 
  T_{\rm b} = \tau_{\rm eff}^{\rm b} =  \frac{\tau^{\rm b}+(\tau^{\rm b})^2 f_{\rm ic}/f_{\rm vor}}{1+\tau^{\rm b}/f_{\rm vor}},   
\end{equation}
\begin{equation} 
  T_{\rm r} = \tau_{\rm eff}^{\rm r} =  \frac{\tau^{\rm b}/2+(\tau^{\rm b})^2 f_{\rm ic}/(4 f_{\rm vor})}{1+\tau^{\rm b}/(2 f_{\rm vor})}.   
  \label{Eq:mdot_full}
\end{equation}
For given $f_{\rm ic}$ this is now a quadratic system with two
distinct solution branches. It is readily solved by any mathematical
software package, but the solutions are too complex to be given
explicitly here. For a given $f_{\rm ic}$, real-valued roots exist
above a given $T_{\rm b}/T_{\rm r}$ threshold, as illustrated by
Fig.~\ref{Fig:mdot_theor}. The blue dashed curves in
Fig.~\ref{Fig:mdot_theor} further show that for physically reasonable
values $f_{\rm ic} \la 0.1$, the first solution branch is
characterized by simply a small correction factor to the previous
expression (eqn.~\ref{Eq:mdot_void}) neglecting $f_{\rm ic}$. But as
also seen from the figure (the red dashed-dotted curves), the other
solution branch represents fundamentally different values of mass-loss
corrections. We can understand this by considering the limiting case
of optically thick clumps and a tenuous inter-clump medium, for which
the bridging law above can be approximated with $\tau_{\rm eff}
\approx f_{\rm vor} + f_{\rm ic} \tau$, resulting in
\begin{equation} 
  f_{\rm vor} = T_{\rm r}(2- T_{\rm b}/T_{\rm r}),   
\end{equation} 
\begin{equation} 
  \frac{\tau^{\rm b}}{T_{\rm b}}  =
   \frac{(\dot{M} q)_{\rm vmSEI}}{(\dot{M} q)_{\rm SEI}} = 
   \frac{2(1-T_{\rm r}/T_{\rm b})}{f_{\rm ic}}.
   \label{Eq:mdot_icm} 
\end{equation}
This solution now offers a second possibility, in addition to
eqn.~\ref{Eq:mdot_void}, to recover the smooth wind results as
  $T_{\rm b}/T_{\rm r} \rightarrow 2$, namely by absorption within the
  inter-clump medium. The scaling of mass-loss correction now is $\sim
  1/f_{\rm ic}$, as illustrated by the black dotted curves in
  Fig.~\ref{Fig:mdot_theor}. Physically, this scaling comes from the
  fact that almost all absorption now takes place in the inter-clump
  medium that fills the velocity-gaps between dense clumps, which for
  typical values $f_{\rm ic} \approx 0.01$ can lead to \textit{very}
  large mass-loss corrections.

The two branches thus represent real solution degeneracies in the
two-component clumped models, physically characterized by absorption
in either i) a mixture of optically thin and thick clumps, or ii)
optically thick clumps where the velocity gaps are being filled in by
the inter-clump medium. Calculations using the vmSEI model confirm
that line profiles from the two solution branches are indeed
identical. This essentially implies that when using resonance doublets
as diagnostic tools in a two-component clumped wind, there will always
be two possibilities to reproduce the same line-profile doublet, even
if one invokes additional constraints (either theoretical, or from
alternative diagnostics) about the inter-clump medium.

\subsection{Vorosity mass-loss correction of Si {\sc iv} in B supergiants} 

\citet{Prinja10} derived $T_{\rm b}/T_{\rm r}$ ratios (averaged within
$0.2 \varv_{\infty} \le \varv \le 0.8 \varv_{\infty}$) for a sample of
B-supergiants using the well separated Si {\sc iv} doublet. We here
simply take their derived ratios and apply the formalism developed
above to obtain the vorosity corrections for mass loss times ion
fraction. Since individual values for $T_{\rm b}$ and $T_{\rm r}$ are
not provided by Prinja \& Massa, we do not derive corresponding values
for $f_{\rm vor}$ in this subsection. For the supersonic wind, LDI
simulations typically predict a very tenuous inter-clump medium, on
order $f_{\rm ic} \approx 0.01$, allowing us to use the simplified
eqns.~\ref{Eq:mdot_void} and \ref{Eq:mdot_icm} for the analysis here
(instead of the more complicated full solution to
eqn.~\ref{Eq:mdot_full}). We further neglect stars in the sample with
unphysically derived values, i.e. those few stars with $T_{\rm
  b}/T_{\rm r} > 2$ or $T_{\rm b}/T_{\rm r} <
1$. Fig.~\ref{Fig:mdot_bsuper} shows our result, giving for the first
solution branch a reasonable average factor of $\sim 5$ for the upward
mass loss correction factor. The same analysis for the second solution
branch gives an average scaling $\sim 50/f_{\rm ic}^{0.01}$, with the
inter-clump medium density parameter measured in units of our standard
choice 0.01. The results also reveal a large scatter about the mean,
indicating that the clumping properties of such B supergiants may vary
quite significantly from star to star.

\subsection{Vorosity mass-loss corrections of PV in $\zeta$ Pup and $\lambda$ Cep}

\begin{figure*}
  \begin{minipage}{9.0cm}
    \resizebox{\hsize}{!}  {\includegraphics[angle=90]{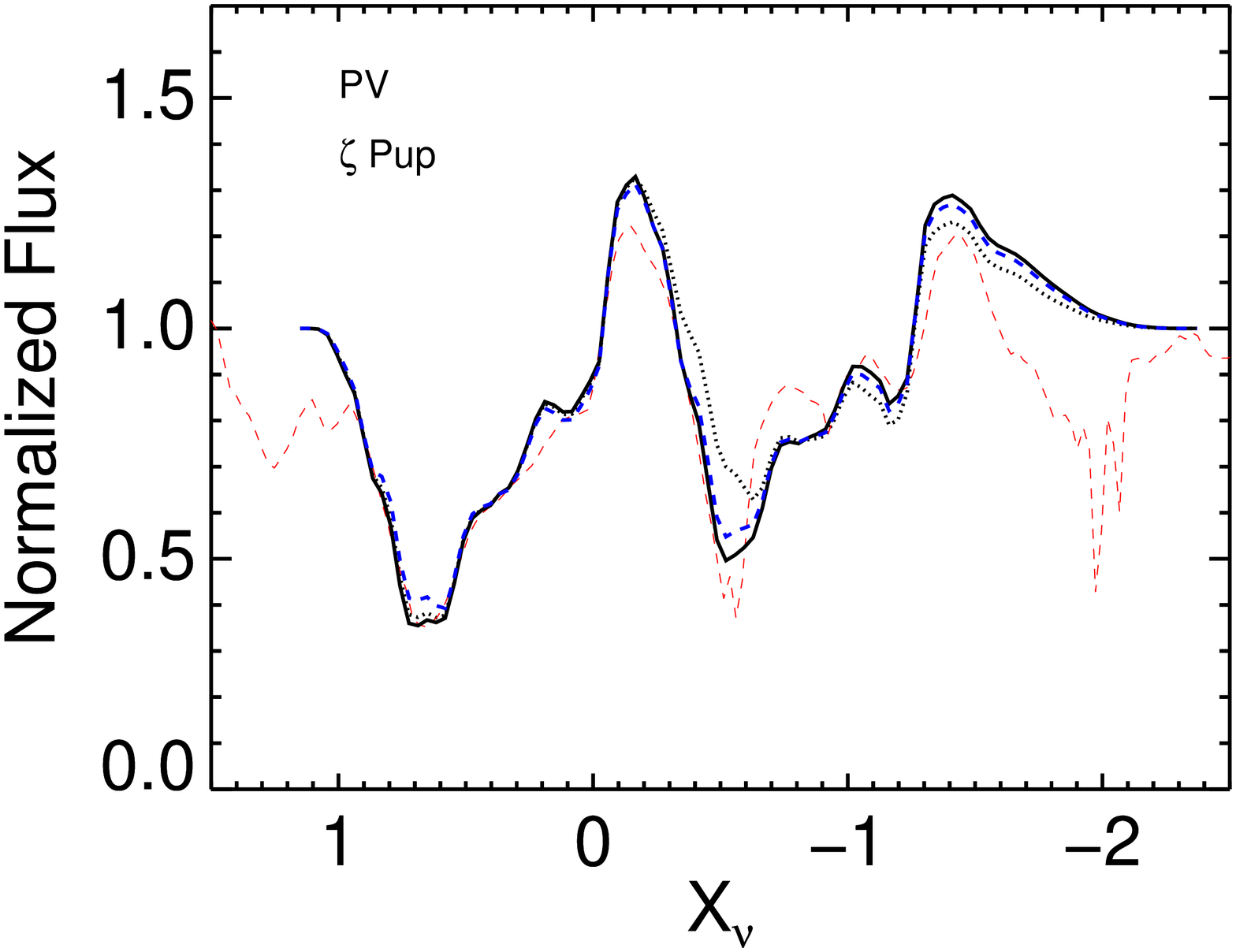}}
    \centering
   \end{minipage}
    \begin{minipage}{9.0cm}
    \resizebox{\hsize}{!}
            {\includegraphics[angle=90]{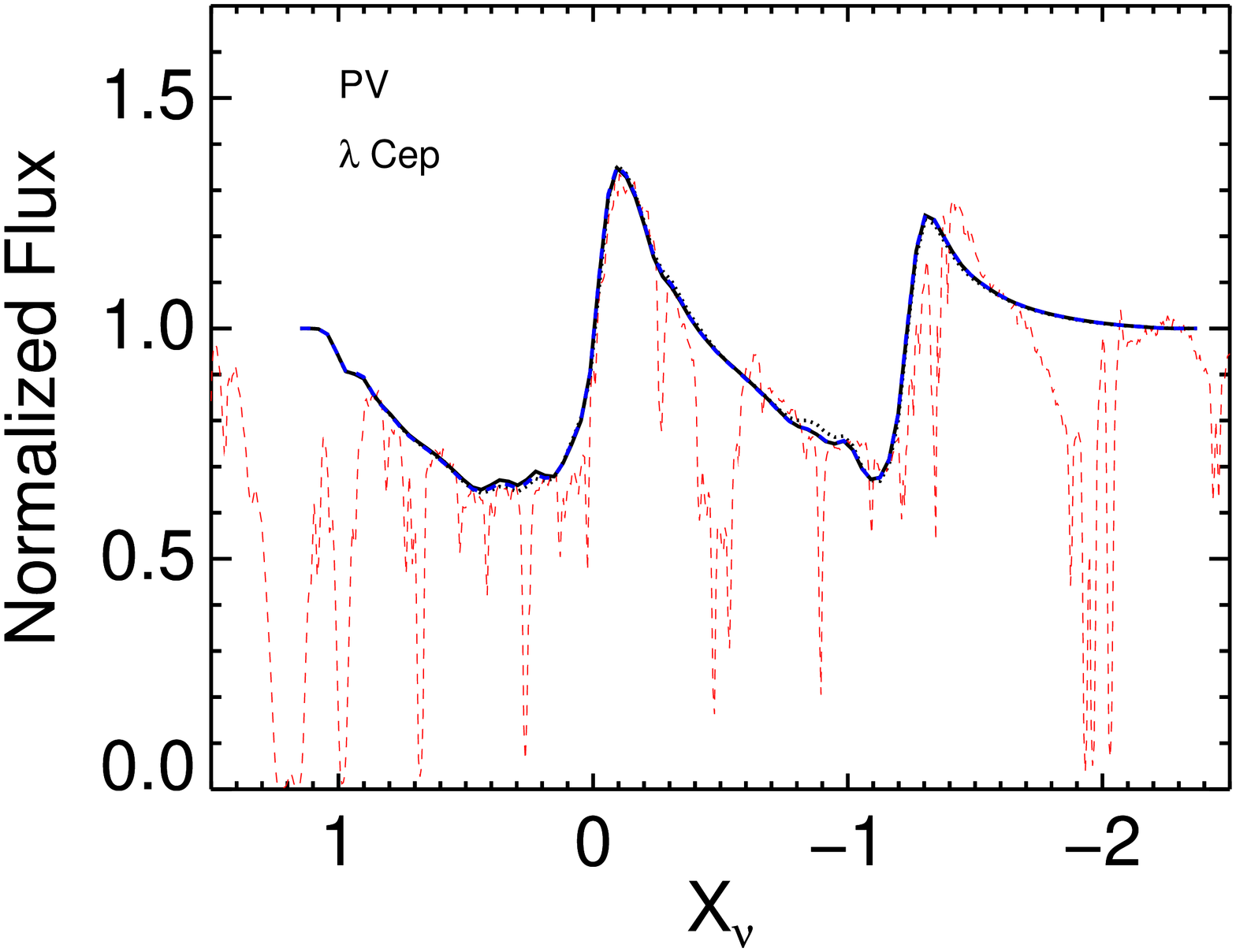}}
            \centering
   \end{minipage}         
  \caption{Observed and modeled PV spectra of $\zeta$ Pup,
    observations from \textit{Copernicus}, and $\lambda$ Cep,
    observations from \textit{FUSE}. Black solid and blue dashed lines
    are vmSEI model fits from the first and second solution branches,
    respectively (see text), the dotted black lines are fits from a
    smooth-wind model, and red dashed lines are the observations.}
  \label{Fig:pv}
\end{figure*}

Using observed spectra from \textit{Copernicus} and \textit{FUSE}
\citep{Fullerton06}, we next perform explicit line-profile fitting of
the unsaturated phosphorus V (PV) resonance doublet at
$\lambda\lambda$1118,1128 in the prototypical Galactic O supergiants
$\zeta$ Pup, O4I(n)f, and $\lambda$ Cep, O6I(n)fp. Using $\beta=0.7$
($\lambda$ Cep) and $\beta = 0.5$ ($\zeta$ Pup) velocity laws
\citep{Fullerton06}\footnote{A velocity field exponent $\beta = 0.5$
  indeed seems somewhat low, but assuming such steep wind acceleration
  actually provides the best fits to the shapes of the PV lines in
  $\zeta$ Pup.} , we first use a normal SEI model with line-strength
$\kappa_0$ and the oscillator strength ratio $f_{\rm b}/f_{\rm r} =
T_{\rm b}/T_{\rm r}$ in 20 discrete velocity bins of
0.05$\varv_\infty$ each as free input parameters. After a simple
$\chi^2$ minimization, the obtained best-fit parameters are translated
to velocity clumping factors and new values for the line-strength
$\kappa_0$ according to the method outlined in previous subsections.

Focusing first on $\zeta$ Pup, the left panel in Fig.~\ref{Fig:pv}
shows profile fits for the two solution branches of the vmSEI model
(black solid and blue dashed lines), and a comparison best-fit
smooth-wind model with a fixed oscillator strength ratio 2 (black
dotted line).  We note from the figure that the vmSEI model fits still
are not perfect; indeed, a completely perfect fit to the observed
\textit{Copernicus} line spectrum would in some velocity bins require
values of $f_{\rm b}/f_{\rm r} = T_{\rm b}/T_{\rm r}$ above 2 or below
1, translating then to unphysical values of the velocity filling
factor. Nonetheless, the vmSEI model is a clear improvement over the
standard smooth-wind SEI fit.  The black solid and blue dashed curves
illustrate the degenerate results from the two solution branches, here
again assuming $f_{\rm ic} = 0.01$. Averaged over velocity bins within
$0.3 \varv \le \varv_\infty \le 0.8\varv$, the first solution results
in a mass-loss times PV ion fraction correction factor for $\zeta$ Pup
of $\langle \dot{M} q \rangle_{\rm vmSEI}/\langle \dot{M} q
\rangle_{\rm SEI} = 6$, and a \textit{normalized velocity filling
factor}
\begin{equation} 
  f_{\rm vel} \equiv \frac{\delta \varv}{\delta \varv + \Delta \varv} = \frac{f_{\rm vor}}{1 +f_{\rm vor}},  
\end{equation} 
of $\langle f_{\rm vel} \rangle = 0.65$. The second solution displayed
in the figure, however, has a \textit{much} higher 
$\langle \dot{M} q \rangle_{\rm vmSEI}/\langle \dot{M} q
\rangle_{\rm SEI} = 60$, and correspondingly a lower $\langle f_{\rm
  vel} \rangle = 0.2$. As discussed in the previous subsection, the
mass-loss corrections from such second branch solutions are further 
scalable in $f_{\rm ic}$, here according to 
$\langle \dot{M} q \rangle_{\rm vmSEI}/\langle \dot{M} q
\rangle_{\rm SEI} \approx 60/f_{\rm ic}^{0.01}$.

For $\lambda$ Cep (the right panel in Fig.~\ref{Fig:pv}), the first
solution branch gives a very modest 20 \% upward correction in mass
loss times ion fraction, accompanied by a high $\langle f_{\rm vel}
\rangle = 0.81$. Note further that the best-fit here is almost
indistinguishable from that using the SEI smooth-wind model with a
fixed $f_{\rm b}/f_{\rm r} = 2$; this is because the line-strength
ratio of the PV blue to red components in $\lambda$ Cep corresponds
almost to a factor of two in optical depth, which leaves essentially
no room for velocity-porosity on the first solution branch (see
Fig.~\ref{Fig:mdot_theor}). On the other hand, using the second
solution tree again results in equally good fits (see the dashed blue
line), and in large mass-loss corrections $\langle \dot{M} q
\rangle_{\rm vmSEI}/\langle \dot{M} q \rangle_{\rm SEI} = 90/f_{\rm
  ic}^{0.01}$, as well as a very low velocity filling factor $\langle
f_{\rm vel} \rangle = 0.04$ for the standard case $f_{\rm ic} = 0.01$.

Physically, these results again reflect the fact that in a
two-component clumped stellar wind, there are two possibilities of
obtaining the same line profile: either by absorption in moderately
optically thick clumps (with clump optical depths reflected in the
observed line-strength ratio), or by filling in the velocity-gaps
between optically thick clumps (which shifts the line-strength ratio
-- which \textit{always} is unity for such optically thick clumps --
to that observed).

It is tempting here to argue that the first solution branch is the
physically more viable, since the inter-clump densities $f_{\rm ic}
\sim 0.01$ typical of LDI wind simulations otherwise would lead to
very large, $\sim 100$, upward corrections in mass loss. As shown
above, applying the first solution for $\zeta$ Pup leads to a factor
of $\sim$\,6 in upward correction of the smooth wind mass-loss rate
times PV ion fraction. Applying this to the PV rate obtained by
\citet{Fullerton06} results in $\langle q \rangle_{\rm PV} \dot{M}
\approx 2.6 \times 10^{-6} \rm \, M_\odot/yr $, which for $\langle q
\rangle_{\rm PV} \approx 0.5-1$, as predicted by present day NLTE
atmosphere codes like {\sc fastwind} \citep{Puls05}, gives a rate in
good agreement with other recent mass loss determinations of this star
\citep{Najarro11, Bouret12, Surlan13, Herve13, Cohen14}.

But as discussed above, for $\lambda$ Cep this solution leads to only
a modest 20 \% upward correction, which (again using the results of
\citealt{Fullerton06}) yields $\langle q \rangle_{\rm PV} \dot{M}
\approx 0.3 \times 10^{-6} \rm \, M_\odot/yr $.  Since $\langle q
\rangle_{\rm PV} \approx 0.5-1$ is predicted also for this star, this
would imply a very low mass-loss rate, a factor of several lower than
that derived from similar velocity-porosity models by
\citet{Sundqvist11} (who essentially ignored the additional
information contained in the doublet-\textit{ratio}) and by
\citet{Surlan13}.
Indeed, the rate obtained by the latter authors corresponds to the
second solution branch found in this paper, which for their very high
assumed $f_{\rm ic} = 0.15$ gives a correction factor $\sim$\,5 for
$\lambda$ Cep and $\langle q \rangle_{\rm PV} \dot{M} \approx 1.4
\times 10^{-6} \rm \, M_\odot/yr $ . This is in good agreement with
the $\dot{M} = 1.6 \times 10^{-6} \rm \, M_\odot/yr $ obtained from
the independent models by \citet{Surlan13}.  (Actually, also the rate
these authors obtain for $\zeta$ Pup corresponds to the second branch
solution; their $\dot{M} = 2.5 \times 10^{-6} \rm \, M_\odot/yr$ for
$f_{\rm ic} = 0.15$ again agrees well with the rate derived in the
analysis above when assuming the second branch solution.) In
Sect.~\ref{discussion}, we further discuss consequences of these
severe degeneracies when empirically deriving mass-loss rates from UV
wind lines.

\section{Application II: Line-driven wind theory}
\label{lines}

Having analyzed in detail how velocity-porosity affects UV spectral
line diagnostics, we next examine the related question of how such
vorosity might affect the line force driving the outflows of hot,
massive stars.

In principle, this effect should be naturally contained within
time-dependent, non-Sobolev simulations of the line-deshadowing
instability. However, many uncertainties regarding, e.g., triggering
of LDI structure \citep{Sundqvist13}, the treatment of the diffuse
force \citep{Owocki99}, and multi-dimensional effects
\citep{Dessart03} are present in such \textit{ab-initio} structured
wind models. Of particular relevance for the study here, is that the
current generation of LDI simulations seems not to properly resolve
the internal clump velocity structures, leading to an overprediction
of the clump velocity spans $\delta \varv$ and to less vorosity than
generally needed to reproduce observations \citep{Sundqvist10,
  Sundqvist11}. Considering these difficulties in creating structured
wind models from first principles, we in this section take an
alternative approach and examine how vorosity might affect the
\textit{global} wind properties mass-loss rate and terminal speed, by
implementing the effective opacity formalism developed above into the
standard, Sobolev-based, theory of line-driven winds.

\subsection{Basic CAK theory}

Let us begin by very briefly review the standard CAK \citep{Castor75}
theory for line driving, cast here in the \citet{Gayley95} formalism
(see also \citealt{Owocki04b}, for a detailed derivation). For
dimensionless line strength $q = \kappa \varv_{\rm th}/(\kappa_{\rm e}
c)$ of the line-center mass absorption coefficient $\kappa$ in units
of the electron scattering opacity $\kappa_{\rm e}$, the CAK model
assumes a power-law distribution of driving lines,
\begin{equation} 
   q \frac{dN}{dq} = \frac{1}{\Gamma(\alpha)} \left( \frac{q}{\bar{Q}} \right)^{\alpha-1}, 
  \label{Eq:CAK_dist}
\end{equation} 
where $\bar{Q}$ is the \citet{Gayley95} line normalization, $\alpha$
the CAK-power index, and $\Gamma$ the gamma function.

For a single line of strength $q$, the line force can be written in
terms of the electron scattering acceleration $g_{\rm e} = \kappa_{\rm
  e} F/c$, with radiative flux $F$, as
\begin{equation} 
  g_{\rm q} = q g_{\rm e} w_{\nu_0} \frac{1-e^{-q t_1}}{q t_1},   
\end{equation} 
where $t_1 = \kappa_{\rm e} \rho c/(d \varv/dr)$ is the radial Sobolev
optical depth for a line with $q =1$, and $w_{\nu_0} \equiv \nu_0
L_{\nu}/L$ weights the placement of the line within the luminosity
spectrum $L_{\nu}$. As usual in CAK theory we here assume a
distribution of driving lines inversely proportional to frequency
about the flux maximum, so that $w_{\nu_0} \approx 1$. The total CAK
line force is then obtained by integrating this single line force over
the number distribution eqn.~\ref{Eq:CAK_dist},
\begin{equation}
  g_{\rm cak} = \int_0^\infty g_{\rm q} \frac{dN}{dq} dq = 
  \frac{\bar{Q}g_{\rm e}}{(1-\alpha)(t_1 \bar{Q})^\alpha}.   
  \label{Eq:CAK_tot} 
\end{equation}
%

\subsection{Vorosity correction to CAK line force}

To correct this for vorosity, we apply the formalism developed in
previous sections, and now write the effective opacity of
line-strength $q$ for clump optical depth $\tau_{\rm cl} = t_{\rm
  q}/f_{\rm vor}$ as
\begin{equation}
  \frac{q_{\rm eff}}{q} = \frac{1 + \tau_{\rm cl} f_{\rm ic}}{1 + \tau_{\rm cl}}.   
\end{equation} 
Anticipating our following results, we re-write this in terms of the
\textit{normalized velocity filling factor} $f_{\rm vel}$ (see
previous section), and further neglect a tenuous inter-clump medium
$f_{\rm ic} \ll 1$ on the driving line force, obtaining after
re-arranging:
\begin{equation} 
  q_{\rm eff} t_1 \approx \frac{f_{\rm vel} q t_1}{f_{\rm vel}+(1-f_{\rm vel})qt_1}. 
\end{equation} 
Applying $q_{\rm eff}$ instead of $q$ in the second expression of
eqn.~\ref{Eq:CAK_tot} now gives the vorosity-modified CAK line force.

The evaluation can be most conveniently carried out in terms of a
correction factor to the standard CAK force,
\begin{equation} 
   \frac{g_{\rm vor}}{g_{\rm cak}} = \frac{1-\alpha}{\Gamma(\alpha)} \int_0^\infty \left[1-\rm Exp \it \left(-\frac{f_{\rm vel} x}{f_{\rm vel}+(1-f_{\rm vel})x}\right) \right] x^{\alpha-2} dx, 
  \label{Eq:cak_vor}
\end{equation}
where the integration dummy variable $x = q t_1$.  

\begin{figure}
              {\includegraphics[width=6.5cm]{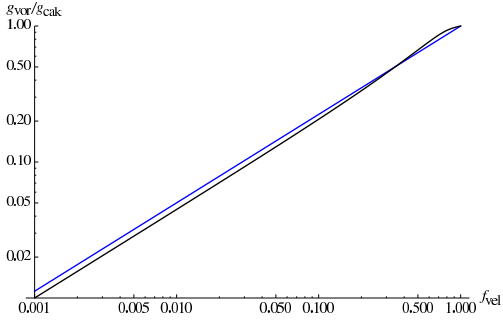}}
  \centering
  \caption{$g_{\rm vor}/g_{\rm cam}$ line forces as functions of $f_{\rm vor}$, for 
  full numerical integration of the vorosity-modified CAK force eqn.~\ref{Eq:cak_vor} 
  (black) and the simple scaling relation eqn.~\ref{Eq:cak_fvor} (blue), for 
  $\alpha  = 2/3$.}
  \label{Fig:cak_fvor}
\end{figure}

Numerical evaluation of eqn.~\ref{Eq:cak_vor} shows this has the
remarkable simple scaling
\begin{equation} 
  \frac{g_{\rm vor}}{g_{\rm cak}} \approx \left(f_{\rm vel}\right)^\alpha,    
  \label{Eq:cak_fvor}
\end{equation} 
as shown by Fig.~\ref{Fig:cak_fvor}, which compares this 
simple approximation to a full integration of 
eqn.~\ref{Eq:cak_vor} for $\alpha = 2/3$. 

Eqn.~\ref{Eq:cak_fvor} thus provides a physically well motivated
rationale for studying the global influence of vorosity on line driven
winds. For clumping close to the wind base, however, the assumption of
neglecting the inter-clump medium's contribution to the line force
might be questionable. But in the absence of good theoretical
predictions or reliable diagnostic results for this wind region, we
retain throughout this section the assumption of a line force
dominated by the dense clumps, also in models involving vorosity in
near photospheric layers. This allows us to study the basic physical
effects and to derive simple scaling relations.

\subsection{Analytic scaling relations for vorosity effect on global wind properties} 

To provide such an analytic rationale for the effects of vorosity on
the global wind parameters mass-loss rate and terminal speed, we
consider the steady-state equation of motion for a wind driven by a
point source of line radiation, in spherical symmetry and in the zero
sound speed limit,
%
\begin{equation} 
 \varv \frac{d \varv}{dr}  = -\frac{GM(1-\Gamma_{\rm e})}{r^2} + g_{\rm cak}, 
\end{equation} 
where $GM(1-\Gamma_{\rm e})/r^2$ is the effective gravitational
acceleration for Eddington parameter $\Gamma_{\rm e} = \kappa_{\rm
  e}L/(4 \pi G M c)$, and where the radiative line acceleration is
assumed to be accurately given by CAK theory. We next introduce the
gravitationally scaled inertial acceleration,
\begin{equation} 
  y' \equiv \frac{r^2 \varv d \varv/dr}{GM(1-\Gamma_{\rm e})}, 
\end{equation} 
and note that for an inverse radius coordinate $x \equiv 1-
R_\star/r$, $y' = dy/dx$ with $y = \varv^2/\varv_{\rm esc}^2$ for
effective escape speed $\varv_{\rm esc}$ reduced by the electron
scattering term. This allows us to write the equation of motion in the
dimensionless form
\begin{equation} 
  y' = -1 + C(w')^\alpha, 
  \label{Eq:eom_cak} 
\end{equation} 
with the constant 
\begin{equation} 
  C \equiv \frac{1}{1-\alpha} \left( \frac{L}{\dot{M}c^2} \right)^\alpha 
  \left( \frac{\bar{Q} \Gamma_{\rm e}}{1-\Gamma_{\rm e}} \right)^{1-\alpha}.  
  \label{Eq:cak_c} 
\end{equation} 
Solving eqn. \ref{Eq:eom_cak} for the tangential intersection between
line force and inertia plus gravity now gives the CAK critical
solution for the \textit{maximal} mass-loss rate and wind velocity law
$\varv \propto \varv_{\rm esc} \sqrt{x}$ \citep[see, e.g.,][for two
  alternative derivations]{Kudritzki89, Owocki04b}. More generally
though, this point-source model should be corrected for the finite
extent of the stellar disc, where
\begin{equation} 
  f_{\rm d}(r) = \frac{(1+\sigma)^{1+\alpha}-(1+\sigma \mu_\star^2)^{1+\alpha}}
  {(1+\alpha)\sigma(1+\alpha)^\alpha(1-\mu_\star^2)}
\end{equation} 
for $\mu_\star^2 = 1-R_\star^2/r^2$ and $\sigma = d \ln \varv/ d \ln r
-1$ is the finite-disc correction factor \citep{Pauldrach86, Friend86}
to the CAK line force. Since $f_{\rm d}$ increases outwards from $r =
R_\star$ (to a certain maximum, typically located at $r \sim 1.5
R_\star$), the stellar surface now represents a nozzle (``throat'')
from which it is hardest to accelerate the material. This fixes the
maximal allowed finite-disc mass loss to
\begin{equation} 
  \dot{M}_{\rm cak}^{\rm fd} =   \frac{L}{c^2} \frac{\alpha}{(1+\alpha)^{1/\alpha}(1-\alpha)}
  \left( \frac{\bar{Q} \Gamma_{\rm e}}{1-\Gamma_{\rm e}} \right)^{1/\alpha-1}, 
  \label{Eq:mdot_cak}
\end{equation} 
which is a factor $1/(1+\alpha)^{1/\alpha}$ lower than the
point-source rate. Since the mass loss in such finite-disc models
quite generally is set close to the wind base, vorosity starting well
above the stellar surface should not affect this scaling. However, if
vorosity is important also at low wind radii, the constant setting the
maximal mass-loss rate will involve an additional factor, $C \sim
1/f_{\rm vel}^{\alpha}$. Due to the scaling $C \sim 1/\dot{M}^\alpha$
then (eqn.~\ref{Eq:cak_c}), the net effect of velocity-porosity is
thus to decrease the mass-loss rate by simply a factor $f_{\rm
  vel}(r_{\rm cp})$, where the velocity filling factor is evaluated at
the ``critical point'' determining this rate.

The wind terminal speed, on the other hand, can be affected also in
cases of a vorosity onset radius above this critical point.
Considering eqn.~\ref{Eq:eom_cak} for a sudden onset of vorosity in
the supersonic $y' \gg 1$ regime, gives the scaling $\varv/\varv_{\rm
  esc} \propto f_{\rm vel}^{\alpha/(2-2\alpha)}$, which for a standard
$\alpha = 2/3$ results in a linear relation $\varv \propto f_{\rm
  vel}$. While the quantitative speed reduction from such outer wind
vorosity, of course, will depend on the exact onset radius, this
simple scaling illustrates that the effect on the terminal speed can
be quite significant.
   
\subsection{Hydrodynamical wind models with vorosity modified CAK line force}

\begin{figure}
  \resizebox{\hsize}{!}
              {\includegraphics[angle=90]{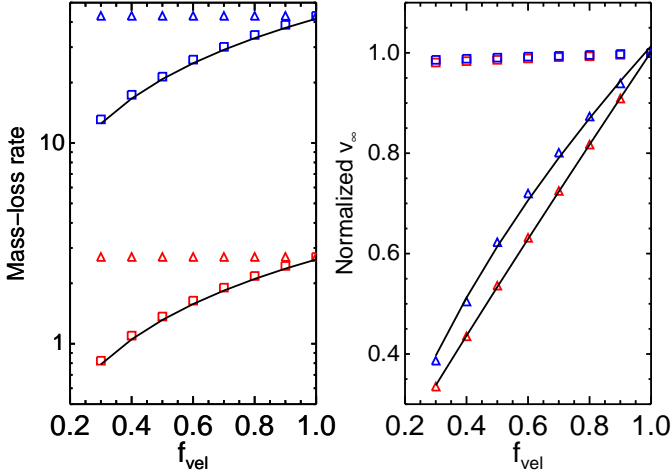}}
  \centering
  \caption{Mass-loss rates in units of $10^{-6} \, \rm M_\odot/yr$ (on
    a logarithmic scale) and terminal speeds in units of
    $\varv_{\infty}$ in standard, non-vorous finite-disc models. The
    triangles and squares denote hydrodynamical calculations
    (described in text) with vorosity onset radii $r_{\rm vor} = 1.3
    \, R_\star > r_{\rm cp}$ (triangles) and $r_{\rm vor} = 1.01 \,
    R_\star < r_{\rm cp}$ (squares). Blue symbols assume $\alpha=0.5$
    and red symbols $\alpha=0.65$. The black solid lines then use the
    corresponding analytic formulae described in text to predict the
    mass-loss rate and terminal speed scalings.}
  \label{Fig:mdot_fvor}
\end{figure}

\begin{table}
    \centering
    \caption{Input parameters used in the hydrodynamical wind simulations described in the text.}
        \begin{tabular}{ l l l}
        Parameter & Symbol & Value  \\     
        \hline \hline 
        Luminosity & $L/L_\odot$ & $8.0 \times 10^5$   \\
        Mass & $M/M_\odot$ & 50   \\
        Radius & $R_\star/R_\odot$ & 20   \\
        Sound speed & $a$ & 23\,km/s   \\
        CAK power-index & $\alpha$ & 0.65, 0.5  \\
	Line normalization & $\bar{Q}$ &  2000  \\
	Electron scattering & &   \\
	mass absorption & $\kappa_{\rm e}$ & 0.345\,$\rm cm^2/g$  \\
                \hline
        \end{tabular}
    \label{Tab:hyd_params}
\end{table}

To back up the analytic scaling results above, we next examine effects
on mass loss and terminal speed by numerically solving the
hydrodynamic conservation equations of mass, momentum, and energy,
using the vorosity-corrected CAK line force just developed.  This line
force is implemented into the hydrodynamics code VH-1 (developed by
J. Blondin and collaborators) according to:
\begin{equation} 
  g_{\rm line}(r) = g_{\rm cak}(r) \  f_{\rm d}(r) \ \left( f_{\rm vel}(r) \right)^\alpha,      
\end{equation} 
where we assume a normalized velocity filling factor
%
\begin{equation} 
  f_{\rm vel}(r) = f_{\rm  vel}^{\infty} + (1-f_{\rm vel}^{\infty} ) \rm Exp \it (-10^3(r/r_{\rm cl}-1)); \ \ r \ge r_{\rm cl},    
  \label{Eq:fvor_vh1}
\end{equation} 
in order to allow for a smooth transition above a vorosity onset
radius $r_{\rm cl}$. To facilitate comparison with the sudden onset of
vorosity assumed when deriving the analytic scaling relations, we have
inserted a factor $10^3$ in the exponential term in
eqn.~\ref{Eq:fvor_vh1}, which ensures that the terminal velocity
filling factor $f_{\rm vel}^{\infty}$ is reached quickly after $r_{\rm
  cl}$. Below $r_{\rm cl}$, $f_{\rm vel}$ is simply set to
unity. Using this line force and assuming for simplicity an isothermal
wind with sound speed $a = 23$\,km/s, the hydrodynamical conservation
equations are evolved until a stable steady-state wind solution is
reached.

For the parameters in Table 2, typical of an early O supergiant in the
Galaxy, Fig.~\ref{Fig:mdot_fvor} compares mass-loss rates in these
numerical models with those predicted by the analytic scaling
\begin{equation}  
  \dot{M}_{\rm vor} = \dot{M}_{\rm cak}^{\rm fd} \ f_{\rm a}  \ f_{\rm vel}(r_{\rm cp}), 
\end{equation} 
where $f_{\rm vel}(r_{\rm cp})$ is the normalized velocity filling
factor at the critical point determining the mass-loss rate, and where
we have also corrected for the finite sound speed $a$ according to the
perturbation analysis by \citet{Owocki04b} \citep[see also Appendix A
  of][]{Owocki04c}:
\begin{equation}
	f_{\rm a} = 1+\frac{4 \sqrt{1-\alpha}}{\alpha} \frac{a}{\varv_{\rm esc}}. 
\end{equation} 	
The left panel of Fig.~\ref{Fig:mdot_fvor} reveals very good agreement
between the numerical models and this simple analytic scaling formula
(within 2-3 \%), for both investigated values of the CAK power index
$\alpha$. We note in particular how, indeed, the mass-loss rate is not
affected by vorosity with an onset radius above the critical point,
which in these finite-disc hydrodynamical simulations lies only a few
percent above the stellar surface.

The right panel of Fig.~\ref{Fig:mdot_fvor} shows the reduction in
$\varv_\infty$, and demonstrates the clear anti-correlation between
mass loss and terminal speed. When vorosity is important below the
critical point, the terminal speed is not affected since the wind then
has adjusted to the reduced line force by lowering the mass
loading. On the other hand, when vorosity is turned on above this
critical point, the wind reacts to the lower line force in the outer
parts by reducing its terminal speed by an amount that follows closely
the analytic scaling $\varv_\infty \propto (f_{\rm
  vel})^{\alpha/(2-2\alpha)}$, as illustrated by the black solid lines
in the right panel of the figure.

\subsection{Ionization correction} 

The computations above assume the line driving parameters $\bar{Q}$
and $\alpha$ are constant throughout the wind. To account for
potential effects on the wind driving from a radially varying
ionization balance, \citet{Abbott82} introduced another correction
factor to the CAK line force, which he took to be $\propto (n_{\rm
  e}^{11}/W)^\delta$ for electron density $n_{\rm e}$ measured in
units of $10^{11}/\rm cm^3$, geometric dilution factor $W$, and
ionization power index $\delta$.  Since $n_{\rm e} \propto \rho$, this
leads to a new scaling of the line force $\sim 1/\rho^{\alpha-\delta}$
(see eqn.~\ref{Eq:CAK_tot}), and the corresponding scaling relations
for $\dot{M}$ are affected only by a different exponent $\alpha_{\rm
  eff} = \alpha - \delta$ (though the absolute values for the
predicted rates may change by some additional factors of order unity,
see e.g.  \citealt{Pauldrach86}). But in a clumped stellar wind with
negligible inter-clump medium, the electron density has to be
evaluated inside the dense clumps, whereby $n_{\rm e} \propto \langle
\rho \rangle/f_{\rm vol}$ and another factor of $(f_{\rm vol})^\delta$
enters the line force expression. Inserting this into the analysis
above then leads to an \textit{upward} correction in the CAK mass-loss
rate, by a factor $(1/f_{\rm vol})^{\delta}$.

We note here these two competing effects from clumping on the line
force; whereas the vorosity-associated reduced line force can lead to
a \textit{lower} mass-loss rate, the shift in ionization balance
stemming from the clumped wind leads to an \textit{increase} in this
rate for typical values of $\delta$. In O stars with $\delta \approx
0.1$ \citep[e.g.,][]{Pauldrach86}, the latter results in an upward
corrected rate by approximately 25 \%, assuming a typical volume
filling factor $f_{\rm vol} \approx 0.1$. Physically, this results
from the increased amount of recombination in such clumped models,
which drives the ionization balance toward lower ion stages with more
efficient driving lines \citep[see also][]{Muijres11}.

\section{Summary and Conclusions}
\label{discussion}

We have developed and benchmarked an effective-opacity formalism for
line (and continuum) radiative transfer in accelerating two-component
media of (almost) arbitrary density contrasts and clump optical
depths. The formalism gives results consistent with our previous, more
elaborate models \citep{Sundqvist10, Sundqvist11}, but is simple
enough that it can be readily included in the already existing NLTE
radiative transfer codes normally used for quantitative modeling and
analysis of spectra from hot stars with winds. In addition to the
clump volume filling factor $f_{\rm vol}$, which enters also standard
descriptions assuming optically thin clumps, the formalism here is
based on two further parameters: $f_{\rm ic}$, defined as the ratio of
the inter-clump density to the mean density, and $\tau_{\rm cl}$, the
clump optical depth. Of course, the method can also be used for the
case of a negligible inter-clump medium, by simply setting $f_{\rm ic}
= 0$. A crucial point is the calculation of $\tau_{\rm cl}$, which for
\textit{continuum} transfer depends on the porosity length $h$, but
for \textit{line} transfer on the velocity clumping factor $f_{\rm
  vor}$ (see Sect. 2). This difference reflects the physics of the
additional leakage of light associated with optically thick clumps in
an accelerating supersonic medium, which for the continuum is set by
\textit{spatial} porosity but for lines by porosity in
\textit{velocity space} (a.k.a. velocity-porosity, or vorosity).

The effective-opacity law for spectral lines is then incorporated into
a vorosity-modified Sobolev with exact integration (vmSEI) method, and
used to analyze unsaturated UV wind resonance line doublets. For a
given inter-clump density $f_{\rm ic}$, an analytic investigation
shows that in clumped two-component winds, two solutions exist that
give identical synthetic line profile doublets. For a given
profile-strength ratio between the two individual line components, the
two solution branches correspond physically to i) absorption within
moderately optically thick clumps with $\tau_{\rm cl}$ determined by
the blue-to-red profile-strength ratio, and ii) absorption within
optically thick clumps ($\tau_{\rm cl} \gg 1$), and the observed
profile-strength ratio reproduced by infilling absorption in the
inter-clump medium. Direct applications to SiIV in the B supergiant
sample of \citet{Prinja10} and to PV in the O supergiants $\zeta$ Pup
and $\lambda$ Cep demonstrate this severe solution dichotomy.  For the
B supergiants and $\zeta$ Pup the physically more realistic first
solution branch gives reasonable mean upward corrections of $\sim$\,5
in mass loss times ion fractions, bringing the PV mass-loss rate for
$\zeta$ Pup into good agreement with other recent studies focusing on
other wavebands than the UV \citep[e.g.,][]{Najarro11, Herve13,
  Cohen14}. The same solution for $\lambda$ Cep, however, gives only a
very modest $\sim 20 \%$ upward correction. This would imply a very
low mass-loss rate of this star, since it seems unlikely that the PV
ion fractions of $\lambda$ Cep and $\zeta$ Pup should be very
different. On the other hand, applying the second solution branch and
assuming a much higher inter-clump density, $\sim 15 \%$ of the mean
density, results in a correction factor $\sim 4-5$ for also this star;
we show that the independent models by \citet{Surlan13} indeed
correspond to this solution. 

In summary, it is very likely that all previous attempts of obtaining
mass-loss rates from fitting UV spectra by means of clumped stellar
wind models -- including our own -- suffer from the uniqueness problem
found in this paper \citep[e.g.,][]{Oskinova07, Sundqvist11,
  Surlan13}. Empirically it seems possible to break these degeneracies
only by a real multi-diagnostic study, in which several diagnostics
are considered simultaneously. In particular, X-ray absorption is a
very promising mass-loss indicator \citep{Cohen10, Cohen11, Herve13,
  Cohen14}, since it has been shown that this diagnostic seems to be
free of most issues associated with wind clumping \citep{Sundqvist12,
  Herve13, Leutenegger13}. Another interesting possibility is to
target stars with very dense winds, like Wolf-Rayet stars or Luminous
Blue Variables in their quiet stage, where effects should be larger
and additional diagnostics are available (for example electron
scattering wings, which are too weak to be of diagnostic value in the
OB-star winds studied here).

Of course, to some extent these degeneracies are artefacts of
present-day diagnostic models, which treat clumping by using a set of
adjustable input parameters rather than computing clumping properties
from first principles. In an ideal situation, one would instead use
simulations of the structured wind to quantitatively predict, e.g.,
vorosity and inter-clump medium properties. However, as discussed in
previous sections, presently such predictions are quantiatively very
uncertain.  For example, the relatively dense inter-clump medium in
the highly supersonic wind indicated by the second branch solutions
discussed above ($f_{\rm ic} > 0.1$), is inconsistent with basic
predictions of the fundamental, inherent instability of line-driving
\citep[e.g.,][]{Owocki88, Feldmeier95, Owocki96, Sundqvist13} that is
the likely cause of clumping in this wind region, and which predicts
much lower inter-clump densities.

We next incorporated the effective-opacity formalism also into the
standard CAK theory of line-driven winds, showing that vorosity leads
to a reduced line force scaling simply with the normalized velocity
filling factor $f_{\rm vel} \equiv f_{\rm vor}/(1+f_{\rm vor})$ as
$f_{\rm vel}^\alpha$, for CAK power index $\alpha$. By analytic and
numerical hydrodynamics calculations, we then derived scaling
relations for the anti-correlated behavior of the global wind
parameters mass-loss rate and terminal speed: For vorosity starting
below the wind ``critical point'', the mass-loss rate is reduced by
factor of $f_{\rm vel}$ but the terminal speed remains unaffected,
whereas for vorosity starting above this critical point the mass-loss
rate is unaffected but the terminal speed reduced by $\varv_\infty
\propto f_{\rm vel}^{\alpha/(2-2\alpha)}$. We finally also provide a
simple correction factor accounting for the expected shift in
ionization in a clumped wind, which for a negligible inter-clump
medium scales as $\dot{M} \propto (1/f_{\rm vol})^\delta$, with
Abbott's ionization parameter $\delta$ ($\approx 0.1$ for a typical O
star wind).

These analytic scalings are qualitatively consistent with the
numerical simulations by \citet{Muijres11}, who modeled the effects of
clumping and porosity in velocity space by using a smooth wind
velocity law and assigning clump length scales\footnote{More precisely,
  they assigned overdensities of clumps $C_{\rm c}$ and mean
  separations $L$, giving clump length scales $l \sim L/C_{\rm
    c}^{1/3}$ and clump velocity spans $\delta \varv \sim (d \varv/ds)
  l$, along path length $s$.}  $l$, and found general trends of higher
mass-loss rates from the shifted ionization balance and lower rates
from the inclusion of velocity porosity. A detailed comparison is not
possible, however, since their Monte-Carlo models only predict the
total wind kinetic energy $\dot{M} \varv_\infty^2$, and so cannot
separate between a change in terminal wind speed and a change in
mass-loss rate.

The upshot from the study here is thus that while vorosity generally
gives an \textit{upward} correction in empirical mass-loss rates
derived from spectral fitting (true for all diagnostics, when compared
to models assuming optically thin clumps), it could also, if there is
substantial vorosity at the wind critical point, cause a
\textit{downward} correction in mass-loss rates predicted by
line-driven wind theory.

Such downward corrections would be consistent with the recent
empirical mass-loss determination of nearby, bright O-stars by
\citet{Cohen14} (using presumably clumping-insensitive X-ray
diagnostics), who find rates that are on average a factor of $\sim$\,3
lower than current theoretical predictions, and also with the many
observational \citep[e.g.,][]{Lepine08, Puls06, Bouret12, Cohen14} and
theoretical \citep{Sundqvist13} findings that strongly indicate
clumping in near photospheric layers. A reduced line force would
further help explain also the long-standing problem of winds from
late-type O main-sequence stars, which seem to be much weaker than
predicted by standard theory \citep[see overview in][]{Puls08}.

Future papers in this series will i) develop more refined theoretical
wind models to account quantitatively for the velocity-porosity
effect, and ii) employ the effective-opacity formalism developed in
this paper in multi-wavelength, multi-diagnostic NLTE studies of hot
star winds in an attempt to break the severe degeneracies discussed
above.

\section{Appendix A} 

Let us consider a two-component ($i=1,2$) mixture described by
homogeneous Markovian statistics, with spatially constant opacities
$\chi_i$ and probabilities $p_{\rm i} = \lambda_{\rm
  i}/(\lambda_1+\lambda_2)$ of at any given point along a ray being in
component $i$, where $\lambda_{\rm i}$ is the mean chord length of
material $i$. In this scenario, the stochastic radiative transfer
equation can be solved analytically for the mean intensity $\langle I
\rangle$ \citep[e.g.,][]{Levermore86}. The book by \citet{Pomraning91}
provides the full derivation; here we merely give the result, along
with a translation of the parameters used by Pomraning and
collaborators to those used in this paper.

The result for the averaged intensity at a distance $s$ along
a ray is
\begin{equation} 
  \langle I \rangle = \Big( \frac{r_{+} - \hat{\sigma}} {r_{+} - r_{-}}\Big) e^{-r_{+} s} + 
  \Big(\frac{ \hat{\sigma}  - r_{-} } {r_{+} - r_{-}}\Big) e^{ -r_{-} s },
  \label{Eq:ifund}
\end{equation}
with 
\begin{equation}
  2r_{\pm} = \langle \chi \rangle + \hat{\sigma} \pm 
  \sqrt{(\langle \chi \rangle - \hat{\sigma})^2 + 4 \beta_{\rm M}},
  \label{Eq:r}
\end{equation}
\begin{equation}
  \hat{\sigma} = p_1 \chi_2 + p_2 \chi_1 + \frac{1}{\lambda_2} + \frac{1}{\lambda_1}, 
  \label{Eq:hat}
\end{equation}
\begin{equation}
  \beta_{\rm M} = (\chi_2 - \chi_1)^2 p_2 p_1,
  \label{Eq:bet}
\end{equation}
and average opacity $\langle \chi \rangle \equiv \chi_1 p_1 + \chi_2 p_2$.

Assuming clumps to be component $1$, we follow the arguments by
\citet{Sundqvist12} and identify $f_{\rm vol} = p_1$, $(1-f_{\rm vol})
= p2$ \citep[see also][]{Pomraning91}, $p_2 \lambda_2 = h$, and
finally $\chi_1 = \chi_{\rm cl}$ and $\chi_2 = \chi_{\rm ic}$. As
demonstrated by the left panel of Fig.~1, these identifications indeed
give perfect agreement for the effective opacity $\chi_{\rm eff} = -
\ln \langle I \rangle / s$ when compared to the numerical 3D box
experiments in Sect. 2.

We note further that also \citet{Pomraning91}, by means of a mean-free
path argument, derives an ``effective opacity'' approximation for their 
two-component model: 
\begin{equation} 
  \chi_{\rm eff}^{\rm Pom} = 
  \frac{\langle \chi \rangle + \chi_{1} \chi_2 \ell_{\rm c}}
  {1+(p_1\chi_2 + p_2 \chi_1)\ell_{\rm c}},  
  \label{Eq:eff_pom} 
\end{equation} 
with correlation length $\ell_{\rm c} = \lambda_1
\lambda_2/(\lambda_1+\lambda_2)$. Translated to the notation used in
this paper, eqn.~\ref{Eq:eff_pom} becomes
\begin{equation} 
  \frac{\chi_{\rm eff}^{\rm Pom}}{\langle \chi \rangle} = 
  \frac{1 + f_{\rm ic} \tau_{\rm cl}/(1-f_{\rm vol})}
       {1 + \tau_{\rm cl}(1 + \frac{f_{\rm vol} \chi_{\rm ic}}
       {(1-f_{\rm vol})\chi_{\rm cl}})},  
       \label{Eq:eff_p} 
\end{equation} 
assuming here $\chi_{\rm ic}/\langle \chi \rangle = f_{\rm ic}$.  In
the limit of $f_{\rm vol} \ll 1$ (for $\chi_{\rm ic} \le \chi_{\rm
  cl}$), eqn.~\ref{Eq:eff_p} simplifies to the effective-opacity law
adopted in Sect. 2:
\begin{equation} 
  \frac{\chi_{\rm eff}^{\rm Pom}}{\langle \chi \rangle} \approx 
  \frac{1 + f_{\rm ic} \tau_{\rm cl}}{1 + \tau_{\rm cl}} = 
  \frac{\chi_{\rm eff}^{\rm SPO}}{\langle \chi \rangle}.  
  \label{Eq:eff_spo} 
\end{equation} 
Testing has shown that this simple bridging law for approximating the
effective opacity actually reproduces the analytic and the numerical
intensity test-calculations in Sect. 2 somewhat \textit{better} than
the more complicated one suggested by Levermore and Pomraning
(eqn.~\ref{Eq:eff_pom}). In particular, eqn.~\ref{Eq:eff_spo} also
represents a very simple extension of an ``intuitive'' law that
corrects the clump+void model (see Sect. 2) by simply adding a tenuous
inter-clump medium with $f_{\rm ic} \ll 1$:
\begin{equation} 
  \frac{\chi_{\rm eff}}{\langle \chi \rangle} \approx
  \frac{1}{1+\tau_{\rm cl}} + f_{\rm ic} = \frac{1+\tau_{\rm cl}f_{\rm
      ic} + f_{\rm ic}}{1+\tau_{\rm cl}}.
  \label{Eq:add} 
\end{equation} 
In contrast to this expression though, the effective opacity law used
in this paper (obtained by simply dropping the alone-standing $f_{\rm
  ic}$ term in the last expression of eqn.~\ref{Eq:add}) also
preserves the ``smooth'' medium limit $\chi_{\rm eff} = \langle \chi
\rangle$ for $f_{\rm ic}=1$.

\begin{acknowledgements}
  JOS gratefully acknowledges support from DFG grant Pu117/8-1.
  J.O.S. and J.P. also acknowledge the help from the International
  Space Science Institute at Bern, Switzerland. We thank John Hillier
  and the anonymous referee for useful comments on the manuscript.
\end{acknowledgements}

\bibliographystyle{aa}
\bibliography{sundqvist_vorIII}

\begin{thebibliography}{55}
\expandafter\ifx\csname natexlab\endcsname\relax\def\natexlab#1{#1}\fi

\bibitem[{{Abbott}(1982)}]{Abbott82}
{Abbott}, D.~C. 1982, \apj, 259, 282

\bibitem[{{Bouret} {et~al.}(2012){Bouret}, {Hillier}, {Lanz}, \&
  {Fullerton}}]{Bouret12}
{Bouret}, J.-C., {Hillier}, D.~J., {Lanz}, T., \& {Fullerton}, A.~W. 2012,
  \aap, 544, A67

\bibitem[{{Castor} {et~al.}(1975){Castor}, {Abbott}, \& {Klein}}]{Castor75}
{Castor}, J.~I., {Abbott}, D.~C., \& {Klein}, R.~I. 1975, \apj, 195, 157

\bibitem[{{Cohen} {et~al.}(2011){Cohen}, {Gagn{\'e}}, {Leutenegger},
  {MacArthur}, {Wollman}, {Sundqvist}, {Fullerton}, \& {Owocki}}]{Cohen11}
{Cohen}, D.~H., {Gagn{\'e}}, M., {Leutenegger}, M.~A., {et~al.} 2011, \mnras,
  415, 3354

\bibitem[{{Cohen} {et~al.}(2010){Cohen}, {Leutenegger}, {Wollman},
  {Zsarg{\'o}}, {Hillier}, {Townsend}, \& {Owocki}}]{Cohen10}
{Cohen}, D.~H., {Leutenegger}, M.~A., {Wollman}, E.~E., {et~al.} 2010, \mnras,
  405, 2391

\bibitem[{{Cohen} {et~al.}(2014){Cohen}, {Wollman}, {Leutenegger}, {Sundqvist},
  {Fullerton}, {Zsarg{\'o}}, \& {Owocki}}]{Cohen14}
{Cohen}, D.~H., {Wollman}, E.~E., {Leutenegger}, M.~A., {et~al.} 2014, \mnras,
  439, 908

\bibitem[{{Dessart} \& {Owocki}(2003)}]{Dessart03}
{Dessart}, L. \& {Owocki}, S.~P. 2003, \aap, 406, L1

\bibitem[{{Feldmeier}(1995)}]{Feldmeier95}
{Feldmeier}, A. 1995, \aap, 299, 523

\bibitem[{{Feldmeier} {et~al.}(2003){Feldmeier}, {Oskinova}, \&
  {Hamann}}]{Feldmeier03}
{Feldmeier}, A., {Oskinova}, L., \& {Hamann}, W.-R. 2003, \aap, 403, 217

\bibitem[{{Friend} \& {Abbott}(1986)}]{Friend86}
{Friend}, D.~B. \& {Abbott}, D.~C. 1986, \apj, 311, 701

\bibitem[{{Fullerton} {et~al.}(2006){Fullerton}, {Massa}, \&
  {Prinja}}]{Fullerton06}
{Fullerton}, A.~W., {Massa}, D.~L., \& {Prinja}, R.~K. 2006, \apj, 637, 1025

\bibitem[{{Gayley}(1995)}]{Gayley95}
{Gayley}, K.~G. 1995, \apj, 454, 410

\bibitem[{{Gr{\"a}fener} {et~al.}(2012){Gr{\"a}fener}, {Owocki}, \&
  {Vink}}]{Grafener12}
{Gr{\"a}fener}, G., {Owocki}, S.~P., \& {Vink}, J.~S. 2012, \aap, 538, A40

\bibitem[{{Gr{\"a}fener} \& {Vink}(2013)}]{Grafener13}
{Gr{\"a}fener}, G. \& {Vink}, J.~S. 2013, \aap, 560, A6

\bibitem[{{Hamann}(1981)}]{Hamann81}
{Hamann}, W.-R. 1981, \aap, 93, 353

\bibitem[{{Hamann} {et~al.}(2008){Hamann}, {Feldmeier}, \&
  {Oskinova}}]{Hamann08}
{Hamann}, W.-R., {Feldmeier}, A., \& {Oskinova}, L.~M., eds. 2008, {Clumping in
  hot-star winds}, Universit\"atsverlag Potsdam 2008

\bibitem[{{Herv{\'e}} {et~al.}(2013){Herv{\'e}}, {Rauw}, \&
  {Naz{\'e}}}]{Herve13}
{Herv{\'e}}, A., {Rauw}, G., \& {Naz{\'e}}, Y. 2013, \aap, 551, A83

\bibitem[{{Hillier}(1991)}]{Hillier91}
{Hillier}, D.~J. 1991, \aap, 247, 455

\bibitem[{{Hillier}(2008)}]{Hillier08}
{Hillier}, D.~J. 2008, in Clumping in Hot-Star Winds, ed. {W.-R.~Hamann,
  A.~Feldmeier, \& L.~M.~Oskinova}, 93--+

\bibitem[{{Kudritzki} {et~al.}(1989){Kudritzki}, {Pauldrach}, {Puls}, \&
  {Abbott}}]{Kudritzki89}
{Kudritzki}, R.-P., {Pauldrach}, A., {Puls}, J., \& {Abbott}, D.~C. 1989, \aap,
  219, 205

\bibitem[{{Lamers} {et~al.}(1987){Lamers}, {Cerruti-Sola}, \&
  {Perinotto}}]{Lamers87}
{Lamers}, H.~J.~G.~L.~M., {Cerruti-Sola}, M., \& {Perinotto}, M. 1987, \apj,
  314, 726

\bibitem[{{L{\'e}pine} \& {Moffat}(2008)}]{Lepine08}
{L{\'e}pine}, S. \& {Moffat}, A.~F.~J. 2008, \aj, 136, 548

\bibitem[{{Leutenegger} {et~al.}(2013){Leutenegger}, {Cohen}, {Sundqvist}, \&
  {Owocki}}]{Leutenegger13}
{Leutenegger}, M.~A., {Cohen}, D.~H., {Sundqvist}, J.~O., \& {Owocki}, S.~P.
  2013, \apj, 770, 80

\bibitem[{{Levermore} {et~al.}(1986){Levermore}, {Pomraning}, {Sanzo}, \&
  {Wong}}]{Levermore86}
{Levermore}, C.~D., {Pomraning}, G.~C., {Sanzo}, D.~L., \& {Wong}, J. 1986,
  Journal of Mathematical Physics, 27, 2526

\bibitem[{{Lucy}(1984)}]{Lucy84}
{Lucy}, L.~B. 1984, \apj, 284, 351

\bibitem[{{Massa} {et~al.}(2008){Massa}, {Prinja}, \&
    {Fullerton}}]{Massa08} {Massa}, D.~L., {Prinja}, R.~K., \&
  {Fullerton}, A.~W. 2008, in Clumping in Hot-Star Winds,
  ed. {W.-R.~Hamann, A.~Feldmeier, \& L.~M.~Oskinova},
  147, Universit\"atsverlag Potsdam 2008

\bibitem[{{Muijres} {et~al.}(2011){Muijres}, {de Koter}, {Vink}, {Krti{\v
  c}ka}, {Kub{\'a}t}, \& {Langer}}]{Muijres11}
{Muijres}, L.~E., {de Koter}, A., {Vink}, J.~S., {et~al.} 2011, \aap, 526, A32

\bibitem[{{Najarro} {et~al.}(2011){Najarro}, {Hanson}, \& {Puls}}]{Najarro11}
{Najarro}, F., {Hanson}, M.~M., \& {Puls}, J. 2011, \aap, 535, A32

\bibitem[{{Oskinova} {et~al.}(2007){Oskinova}, {Hamann}, \&
  {Feldmeier}}]{Oskinova07}
{Oskinova}, L.~M., {Hamann}, W.-R., \& {Feldmeier}, A. 2007, \aap, 476, 1331

\bibitem[{{Owocki}(2004)}]{Owocki04b}
{Owocki}, S. 2004, in EAS Publications Series, Vol.~13, EAS Publications
  Series, ed. M.~{Heydari-Malayeri}, P.~{Stee}, \& J.-P. {Zahn}, 163--250

\bibitem[{{Owocki}(2008)}]{Owocki08}
{Owocki}, S.~P. 2008, in Clumping in Hot-Star Winds, ed. W.-R. {Hamann},
  A.~{Feldmeier}, \& L.~M. {Oskinova}, 121--

\bibitem[{{Owocki} {et~al.}(1988){Owocki}, {Castor}, \& {Rybicki}}]{Owocki88}
{Owocki}, S.~P., {Castor}, J.~I., \& {Rybicki}, G.~B. 1988, \apj, 335, 914

\bibitem[{{Owocki} \& {Cohen}(2006)}]{Owocki06}
{Owocki}, S.~P. \& {Cohen}, D.~H. 2006, \apj, 648, 565

\bibitem[{{Owocki} {et~al.}(2004){Owocki}, {Gayley}, \& {Shaviv}}]{Owocki04}
{Owocki}, S.~P., {Gayley}, K.~G., \& {Shaviv}, N.~J. 2004, \apj, 616, 525

\bibitem[{{Owocki} \& {Puls}(1996)}]{Owocki96}
{Owocki}, S.~P. \& {Puls}, J. 1996, \apj, 462, 894

\bibitem[{{Owocki} \& {Puls}(1999)}]{Owocki99}
{Owocki}, S.~P. \& {Puls}, J. 1999, \apj, 510, 355

\bibitem[{{Owocki} \& {Rybicki}(1984)}]{Owocki84}
{Owocki}, S.~P. \& {Rybicki}, G.~B. 1984, \apj, 284, 337

\bibitem[{{Owocki} \& {Rybicki}(1985)}]{Owocki85}
{Owocki}, S.~P. \& {Rybicki}, G.~B. 1985, \apj, 299, 265

\bibitem[{{Owocki} \& {ud-Doula}(2004)}]{Owocki04c}
{Owocki}, S.~P. \& {ud-Doula}, A. 2004, \apj, 600, 1004

\bibitem[{{Pauldrach} {et~al.}(1986){Pauldrach}, {Puls}, \&
  {Kudritzki}}]{Pauldrach86}
{Pauldrach}, A., {Puls}, J., \& {Kudritzki}, R.~P. 1986, \aap, 164, 86

\bibitem[{{Pomraning}(1991)}]{Pomraning91}
{Pomraning}, G.~C. 1991, {Linear kinetic theory and particle transport in
  stochastic mixtures}, ed. {Pomraning, G.~C.}, New Jersey: World Scientific, c1991, 
Series on advances in mathematics for applied sciences, v. 7

\bibitem[{{Prinja} \& {Massa}(2010)}]{Prinja10}
{Prinja}, R.~K. \& {Massa}, D.~L. 2010, \aap, 521, L55

\bibitem[{{Prinja} \& {Massa}(2013)}]{Prinja13}
{Prinja}, R.~K. \& {Massa}, D.~L. 2013, \aap, 559, A15

\bibitem[{{Puls} {et~al.}(2006){Puls}, {Markova}, {Scuderi}, {Stanghellini},
  {Taranova}, {Burnley}, \& {Howarth}}]{Puls06}
{Puls}, J., {Markova}, N., {Scuderi}, S., {et~al.} 2006, \aap, 454, 625

\bibitem[{{Puls} {et~al.}(1993){Puls}, {Owocki}, \& {Fullerton}}]{Puls93}
{Puls}, J., {Owocki}, S.~P., \& {Fullerton}, A.~W. 1993, \aap, 279, 457

\bibitem[{{Puls} {et~al.}(2005){Puls}, {Urbaneja}, {Venero}, {Repolust},
  {Springmann}, {Jokuthy}, \& {Mokiem}}]{Puls05}
{Puls}, J., {Urbaneja}, M.~A., {Venero}, R., {et~al.} 2005, \aap, 435, 669

\bibitem[{{Puls} {et~al.}(2008){Puls}, {Vink}, \& {Najarro}}]{Puls08}
{Puls}, J., {Vink}, J.~S., \& {Najarro}, F. 2008, \aapr, 16, 209

\bibitem[{{Sundqvist} \& {Owocki}(2013)}]{Sundqvist13}
{Sundqvist}, J.~O. \& {Owocki}, S.~P. 2013, \mnras, 428, 1837

\bibitem[{{Sundqvist} {et~al.}(2012{\natexlab{a}}){Sundqvist}, {Owocki},
  {Cohen}, {Leutenegger}, \& {Townsend}}]{Sundqvist12}
{Sundqvist}, J.~O., {Owocki}, S.~P., {Cohen}, D.~H., {Leutenegger}, M.~A., \&
  {Townsend}, R.~H.~D. 2012{\natexlab{a}}, \mnras, 420, 1553

\bibitem[{{Sundqvist} {et~al.}(2012{\natexlab{b}}){Sundqvist}, {Owocki}, \&
  {Puls}}]{Sundqvist12c}
{Sundqvist}, J.~O., {Owocki}, S.~P., \& {Puls}, J. 2012{\natexlab{b}}, in
  Astronomical Society of the Pacific Conference Series, Vol. 465, Proceedings
  of a Scientific Meeting in Honor of Anthony F. J. Moffat, ed. L.~{Drissen},
  C.~{Rubert}, N.~{St-Louis}, \& A.~F.~J. {Moffat}, 119

\bibitem[{{Sundqvist} {et~al.}(2010){Sundqvist}, {Puls}, \&
  {Feldmeier}}]{Sundqvist10}
{Sundqvist}, J.~O., {Puls}, J., \& {Feldmeier}, A. 2010, \aap, 510, 11

\bibitem[{{Sundqvist} {et~al.}(2011){Sundqvist}, {Puls}, {Feldmeier}, \&
  {Owocki}}]{Sundqvist11}
{Sundqvist}, J.~O., {Puls}, J., {Feldmeier}, A., \& {Owocki}, S.~P. 2011, \aap,
  528, 64

\bibitem[{{{\v S}urlan} {et~al.}(2013){{\v S}urlan}, {Hamann}, {Aret},
  {Kub{\'a}t}, {Oskinova}, \& {Torres}}]{Surlan13}
{{\v S}urlan}, B., {Hamann}, W.-R., {Aret}, A., {et~al.} 2013, \aap, 559, A130

\bibitem[{{{\v S}urlan} {et~al.}(2012){{\v S}urlan}, {Hamann}, {Kub{\'a}t},
  {Oskinova}, \& {Feldmeier}}]{Surlan12}
{{\v S}urlan}, B., {Hamann}, W.-R., {Kub{\'a}t}, J., {Oskinova}, L.~M., \&
  {Feldmeier}, A. 2012, \aap, 541, A37

\bibitem[{{Zsarg{\'o}} {et~al.}(2008){Zsarg{\'o}}, {Hillier}, {Bouret}, {Lanz},
  {Leutenegger}, \& {Cohen}}]{Zsargo08}
{Zsarg{\'o}}, J., {Hillier}, D.~J., {Bouret}, J.-C., {et~al.} 2008, \apjl, 685,
  L149

\end{thebibliography}

\end{document}